\documentclass[12pt]{emulateapj}
\usepackage{txfonts}

\def\ltsima{$\; \buildrel < \over \sim \;$}
\def\simlt{\lower.5ex\hbox{\ltsima}}
\def\gtsima{$\; \buildrel > \over \sim \;$}
\def\simgt{\lower.5ex\hbox{\gtsima}}
  \def\Sec{${}^{\prime\prime}$\llap{.}}

\shorttitle{On hydrogen and helium burning variables} 
\shortauthors{Coppola et al.}

\begin{document}
\title{The Carina project IX: On hydrogen and helium burning variables.}

\author{G. Coppola\altaffilmark{1},
M. Marconi\altaffilmark{1},
P. B. Stetson\altaffilmark{2},
G. Bono\altaffilmark{3,4},
V. F. Braga\altaffilmark{3,4},
V. Ripepi\altaffilmark{1},
M. Dall'Ora\altaffilmark{1},
I. Musella\altaffilmark{1},
R. Buonanno\altaffilmark{3,5},
M. Fabrizio\altaffilmark{5},
I. Ferraro\altaffilmark{4},
G. Fiorentino\altaffilmark{6},
G. Iannicola\altaffilmark{4},
M. Monelli\altaffilmark{7,8},
M. Nonino\altaffilmark{9},
F. Th\'evenin\altaffilmark{10} and
A. R. Walker\altaffilmark{11}
}

\altaffiltext{1}{INAF--Osservatorio Astronomico di Capodimonte, Via Moiariello 16, 
80131 Napoli, Italy; email: coppola@na.astro.it} 
\altaffiltext{2}{National Research Council, 5071 West Saanich Road,
Victoria, BC V9E 2E7, Canada}
\altaffiltext{3}{Dipartimento di Fisica, Universit\'{a} di Roma Tor Vergata, Via della Ricerca Scientifica 1, 00133 Roma, Italy}
\altaffiltext{4}{INAF-Osservatorio Astronomico di Roma, Via Frascati
  33, 00040 Monte Porzio Catone, Italy}
\altaffiltext{5}{INAF-Osservatorio Astronomico di Teramo, via
  M. Maggini, I-64110, Teramo, Italy}
\altaffiltext{6}{INAF-Osservatorio Astronomico di Bologna, via Ranzani 1, 40127, Bologna}
 \altaffiltext{7}{Instituto de Astrof\'{i}sica de Canarias, Calle Via Lactea s/n, E38205 La Laguna, Tenerife, Spain}
\altaffiltext{8}{Departmento de Astrof\'{i}sica, Universidad de La
  Laguna, E38200 La Laguna, Tenerife, Spain}
\altaffiltext{9}{INAF-Osservatorio Astronomico di Trieste, via
  G. B. Tiepolo 11, I-40131 Trieste, Italy}
\altaffiltext{10}{Laboratoire Lagrange, Université Côte d'Azur, Observatoire de la Côte d'Azur, CNRS, 
Blvd de l'Observatoire, CS 34229, 06304 Nice cedex 4, France}
\altaffiltext{11}{Cerro Tololo Inter-American Observatory, National Optical Astronomy Observatory, Casilla 603, La Serena, Chile}

\begin{abstract}
We present new multi-band ($UBVI$) time-series data of helium burning
variables in the Carina dwarf spheroidal galaxy. The current sample 
includes 92 RR Lyrae--six of them are new identifications--and 20
Anomalous Cepheids, one of which is new identification. 
The analysis of the Bailey diagram shows that the luminosity amplitude 
of the first overtone component in double-mode variables is located along
the long--period tail of regular first overtone variables, while the
fundamental component is located along the short--period tale of 
regular fundamental variables. This evidence further supports the 
transitional nature of these objects.   
Moreover, the distribution of Carina double--mode variables in the 
Petersen diagram (P$_1$/P$_0$ vs P$_0$) is similar to metal--poor 
globulars (M15, M68), to the dwarf spheroidal Draco and to the Galactic 
Halo. This suggests that the Carina old stellar population is 
metal-poor and affected by a small spread in metallicity. 
We use trigonometric parallaxes for five field RR Lyrae
stars~\citep{benedict11} 
to provide an independent estimate of the Carina
distance using the observed reddening free Period--Wesenheit 
[PW, ({\it BV})] relation. Theory and observations indicate that 
this diagnostic is independent of metallicity. We found a true 
distance modulus of 
$\mu$=20.01$\pm$0.02 (standard error of the mean) $\pm$0.05 (standard deviation) mag.
We also provided independent estimates of the Carina true distance modulus using 
four predicted PW relations ({\it BV, BI, VI, BVI}) and we found: 
$\mu$=(20.08$\pm$0.007$\pm$0.07) mag,  $\mu$=(20.06$\pm$0.006$\pm$0.06) mag,
$\mu$=(20.07$\pm$0.008$\pm$0.08) mag and
$\mu$=(20.06$\pm$0.006$\pm$0.06) mag.
Finally, we identified more than 100 new SX Phoenicis stars that together with
those already known in the literature (340) make Carina a fundamental laboratory
to constrain the evolutionary and pulsation properties of these transitional
variables.

\end{abstract}


% %%%%%%%%%%%%%%%%%%%%%%%%%%%%%%%%%%%%%%%%%%%%%%%%%%%%%%%%%%%%%%%%%%%%%%%%%%%%%%%%
\section{Introduction}
Dwarf galaxies play a crucial role in several astrophysical and cosmological 
problems. They outnumber giant stellar systems in the nearby universe and 
at low redshift~\citep[see e.g.][]{mateo98, mc2012}, but it seems that the 
number is steadily decreasing at large redshifts \citep{baldry2012,mortlock2013}. 
The current empirical evidence indicates that the 
luminosity function does not peak at low--surface brightness systems at 
redshifts larger than z=2~\citep{weinzirl11, bauer11}. It is clear that in 
this context the age distribution of the old stellar populations in nearby 
dwarf galaxies plays a crucial role in constraining their early formation 
and evolution~\citep{salvadori10}. 
Current empirical evidence indicates that both early-
and late-type dwarf galaxies host stellar populations older than 10 Gyr. 
This evidence applies not only to IC10, the Local Group prototype of star 
forming galaxies~\citep{sanna09}, but also to the recently discovered 
Ultra Faint Dwarf (UFD)  
galaxies~\citep{dallora06,kuehn08,greco08,moretti09,musella09,dallora12,musella12,clementini12,garofalo13,
fabrizio14}. 
Moreover, the comparison between old stellar populations in nearby dwarfs and 
Galactic Globular clusters does not show, at fixed metal content, any striking 
difference concerning the age distribution ~\citep{monelli03,cole07,bono10}. 

The above empirical evidence suggests that low-mass dark matter halos started
assembling baryons at about the same time as the giant 
halos~\citep{springel05, bauer11, duncan14}. 
Detailed constraints on the formation of these systems 
require detailed investigations into their star-formation histories. 
However, this approach needs deep and accurate color-magnitude 
diagrams (CMDs) down to limiting magnitudes fainter than the main-sequence 
turnoff of the old stellar populations~\citep{gallart05,monelli10a}. 
In this context variable stars play a crucial role, since 
they can be easily identified. Moreover, they cover a broad range 
in age, from a few hundred Myr for Classical Cepheids~\citep[e.g.][]{bono05}  
to a few Gyr for Anomalous Cepheids~\citep[e.g.][]{caputo98,marconi04} to 
ages older than 10 Gyr for the RR Lyrae stars~\citep[e.g.][]{ccp91}. 
However, for classical Cepheids we have evidence of tight correlation 
between pulsation and evolutionary age~\citep{mats11, mats13}. 
The same outcome does not apply, as noted by the referee, to the other 
quoted variables. This means that for RR Lyrae stars we can only provide a lower 
limit to their individual ages~\citep{bono11}, while 
for anomalous Cepheids we can only provide an age interval.

The latter two groups have several distinctive features that make 
them solid stellar tracers in resolved dwarf spheroidal galaxies. 
Empirical evidence indicates that all the dSphs that have been 
searched for evolved helium burning variable stars host RR Lyrae 
stars (hereinafter RRLs). It is worth mentioning that they have been identified in all 
stellar systems hosting a stellar population older than 10 Gyr 
and an intermediate horizontal branch (HB) morphology, i.e., systems in which the 
mass distribution along the HB covers the RRL instability 
strip. The stellar systems with ages younger than 10 Gyr typically display 
Galactic thin disk kinematics~\citep{blc2015}. This means 
that we still lack solid empirical evidence of RR Lyrae belonging to 
old thin disk open clusters~\citep{bono11}.
On the other hand, Anomalous Cepheids (hereinafter AC) have been identified 
in stellar systems more metal-poor than about [Fe/H]=--1.6 dex. Theoretical 
predictions indicate that helium burning sequence of intermediate-mass 
stars more metal-rich than the above limit do not cross the so-called 
Cepheid instability strip \citep{stetson14a,castellanideglinnocenti1995}.

It goes without saying that the comparison of pulsation and evolutionary 
properties of evolved helium burning-variables in nearby stellar systems 
allows us to constrain their early formation~\citep{fiorentino12,coppola13,stetson14a,fiorentino15a}. 
The Carina dSph is a very interesting laboratory
for investigating the above issues. In a previous investigation we
compared the period distribution of the central helium burning variable stars in Carina with similar 
distributions in nearby dSphs and in the Large Magellanic 
Cloud and we found that the old stellar populations in these systems
share similar properties~\citep[][hereinafter, Paper VI]{coppola13}. 
On the other hand, the period distribution and the Bailey diagram
(luminosity amplitude
vs period) of ACs show 
significant differences among the above stellar systems. This evidence suggested 
that the properties of intermediate-age stellar populations might be affected 
both by environmental effects and structural parameters. 

In this investigation we move towards a complete census of helium-burning 
variable stars. 
The structure of the paper is the following. In \S~2 we present in detail
the adopted photometric data set together with the light curves of both
RRLs and ACs and their pulsation properties.  In \S~3 we discuss the
properties of RRLs and ACs using the Bailey diagram and the Petersen 
diagram (period ratio vs period). We focussed
our attention on the position of double mode variables and on the sensitivity
of the period ratio on the iron abundance.
In \S~4 we use the optical Period--Wesenheit--Metallicity (PWZ) relation of
RRLs to estimate the Carina true distance modulus. To constrain the possible
occurrence of systematic errors we estimated the Carina distance using the
five field RRLs for which are available accurate estimates of their trigonometric
parallaxes. Independent distance determinations were also provided using predicted
optical PWZ relations. In \S~5 we present preliminary results about SX
Phoenicis variables. Finally, in \S~6 we summarize the results of this
investigation and briefly outline the future development of the Carina project.

%%%%%%%%%%%%%%%%%%%%%%%%%%%%%%%%%%%%%%%%%%%%%%%%%%%%%%%%%%%%%%%%%%%%%%%%%%%%%%%%
\section{Photometric data}\label{sec:data}
The photometric catalog adopted in this investigation is an extension of the
data set discussed by~\citet{coppola13}. This included  
4,474 CCD images in the {\it UBVRI\/} photometric bands
obtained with the CTIO 4m and 1.5m telescopes, the ESO 3.6m NTT,
the MPI/ESO 2.2m telescope, and the 8m ESO VLT during the period 1992 December
through 2008 September. Here we add 2,028 CCD images in the same
photometric bands obtained with the same telescopes (see
Table~\ref{tab_obs}). In the current analysis, the $R$-band data were not included
because the number of measurements is limited and they do not provide a good 
coverage of the pulsation cycle. The seeing had a median value of 1\Sec0, and 
ranged from 0\Sec8 (10-percentile) and 0\Sec9 (25-percentile) to 1\Sec3 
(75-percentile) and 1\Sec6 (90-percentile). 
The reader interested in a detailed discussion of the photometric methodology, 
the absolute photometric calibration, the identification of the variable stars 
and analysis of the time series data is referred to Paper VI and 
to~\cite{stetson14a}.  

The list of variable stars begins with that from~\citet{dallora03}
(hereinafter D03), who merged the prior lists of~\citet{sms86}, and performed 
their own comprehensive investigation of the resulting set of 92
candidate variable stars.  
This existing star list has been augmented by our own search for new
variables within the new, expanded set of 6,992 optical images of Carina.  
In particular, we have identified 92 RRLs. 
Among them 12 pulsate in the first-overtone (FO, RR$_{c}$), 
63 in the fundamental mode (F, RR$_{ab}$),  
nine double-mode pulsator (RR$_{d}$) 
and eight candidate Blazhko variables (Bl). We also identified 
20 ACs, two long-period variables (LPV) 
and 10 geometrical variables (eclipsing binaries, EB, and W Uma). 
The main difference between the current set of variable stars 
and those discussed in Paper VI is that  we have identified 16 new variables. 
Among them 6 are RRLs, one AC, seven EBs and two probable LPVs.  
Moreover, in Paper VI we discovered 14 new variables, but by using the
new data set, we confirm that only nine of these are new
identifications (see Appendix for details).

Recently~\citet{vivas13} (hereinafter VM13) performed a detailed 
investigation of pulsating stars in Carina. They found 38 RRLs, ten AC variables 
and more than 340 new SX Phoenicis (SX Phe). Among them 36 
(30 RRLs and six ACs) are in common with D03 and 
41 (32 RRLs, seven ACs, two EBs) with our new catalog. The nine stars listed 
by VM13 which do not appear in our catalog are listed in Table~\ref{tab_prop_vivas}. These 
variables were not been included in the following analysis of RRLs pulsation 
properties since they are located beyond the Carina tidal radius (see 
\S~\ref{sec:bailey} for a detailed discussion). Concerning SX Phe stars, 
we confirm the variability for 324 out of 340 known pulsators in Carina and we 
identified 101 new variables of this class (see \S~5 for details ). 

In the first column of Table~\ref{tab_coordinates} we  find the
identification number according to S86 and D03.
For the newly detected variables in this paper and in Paper VI (208--237) the D03 running number 
was continued and these stars also marked with an asterisk in this
same table. Columns (2) and (3) list $\alpha$ and 
$\delta$ (J2000.0) coordinates in units of ($hh:mm:ss$) and ($dd:mm:ss$), respectively\footnote{
The astrometry is on the system of the USNO-A2.0 catalog.}, while the last 
five columns give S86, D03, VM13, Paper VI and current individual notes. 
Candidate variable stars for which we could not confirm the 
variability are given in Table~\ref{tab_nvcoordinates}.

The current data set included the {\it BV} images adopted by 
D03 in their investigation of Carina's variable stars. 
We confirm their variability analysis and their variable 
star classification with only a few exceptions.  
According to our extended  
data set the variables V161 does not show clear sign of variability. 
The variable V17 appears to be an EB variable; while the 
variables V22 and V176 seem to pulsate in the fundamental 
(RR$_{ab}$) instead of in the first overtone (RR$_{c}$) mode. The 
variables V31, V61, V77, V126, V127 and V206 were also classified 
by D03 as RR$_{ab}$ variables and according to our new data
they also show the Blazhko effect.
Moreover, the variable V74 was classified 
by D03 as RR$_{c}$ variable, but according to our new data
it is classified as a double-mode pulsator. A more detailed analysis of individual 
variables is given in the Appendix. 

The individual {\it UBVI} measurements for all the variables in our 
sample are listed in
Table~\ref{tab_phot}\footnote{The Table is presented 
in its entirety in the electronic edition of the paper.}. For every variable in our sample the Table 
gives in the first three columns the Heliocentric Julian Date (HJD), 
the $U$-band magnitude and the photometric error. Columns from (4) to (12) 
give the same information, but for the $B$-, $V$-, and $I$-band measurements. 
The total number of phase points in the complete data set depends on the 
photometric band, and they range from 1 to 23 ($U$), from 22 to 203 ($B$), 
from 38 to 278 ($V$) and from 3 to 92 ($I$). The coverage of the pulsation 
cycle is optimal in the $B$ and in the $V$ band, modest in the $I$ band 
and poor in the $U$ band.    
The photometric error of individual measurements depends on the photometric 
band and on seeing conditions. It is also occasionally affected by crowding conditions, but 
it is on average of the order of 0.02 mag. 

Period searches and the phasing of light curves were performed using the
procedure adopted by~\citet{stetson14b}, for the galactic
Globular Cluster M4. The large time baseline ($\sim$20 years) covered by our data 
set allowed us to overcome half- and one-day alias ambiguities 
that can be quite severe in data sets covering limited 
time intervals. The large time interval covered by the current 
data set allowed us to provide more precise periods, epochs of 
maximum light and luminosity amplitudes for the entire set of 
variable stars. In particular, the accuracy of the period estimates 
is related to the period itself and to the time interval covered 
by the observations. They range from (1$\times$$10^{-8}$) to 
(1$\times$$10^{-5}$) days.

The mean optical magnitudes and the amplitudes in the bands with good time sampling ($B$,$V$)
were estimated from fits with a spline under tension.

We already mentioned that the time sampling of both the $U$- and $I$-band 
light curves is either modest or poor. This means less accurate mean
magnitudes and luminosity amplitudes. To overcome this problem  we 
adopted for the $I$-band light curves the method described 
by~\cite{dicriscienzo11} based on using the $V$-band light curve as a template 
and re-scaling the same measurements in amplitude to fit the $I$-band 
light curves. To accomplish this goal we adopted a visual-to-$I$-band 
amplitude ratio of 1.58 $\pm$  0.03, obtained by~\cite{dicriscienzo11}
using the literature $V$ and $I$ light curves of 130 RRLs with
good light curve parameters selected from different Galactic globular
clusters.  

The $U$-band light curves for several variables are poorly sampled, 
and often there is no coverage across minimum and/or maximum 
light. In those cases, we adopted as mean magnitude the median 
of measurements and we do not provide luminosity amplitudes.

Table~\ref{tab_prop} lists from left to right for every variable in 
the current data set the identification, the classification, the epoch 
of maximum light, the period (days), the intensity-averaged [$<U>$, $<B>$, $<V>$, 
$<I>$] mean magnitudes and the A($U$), A($B$), A($V$), A($I$) luminosity 
amplitudes. For some variables the light curves are poorly sampled, 
and as explained above for the $U$ band,  we adopted as 
mean magnitude the median of measurements and do not provide luminosity amplitudes. 
Note that for RR$_{d}$ variables the same observables and 
the period ratios, $P_{0}/P_{1}$, are listed in Table~\ref{tab_prop_rrd}. 
For these stars we did not perform the analysis of pulsational parameters 
in the $U$-band, because the $U$-band light curves are not well
sampled. 

\begin{figure*}
\includegraphics[scale=0.9]{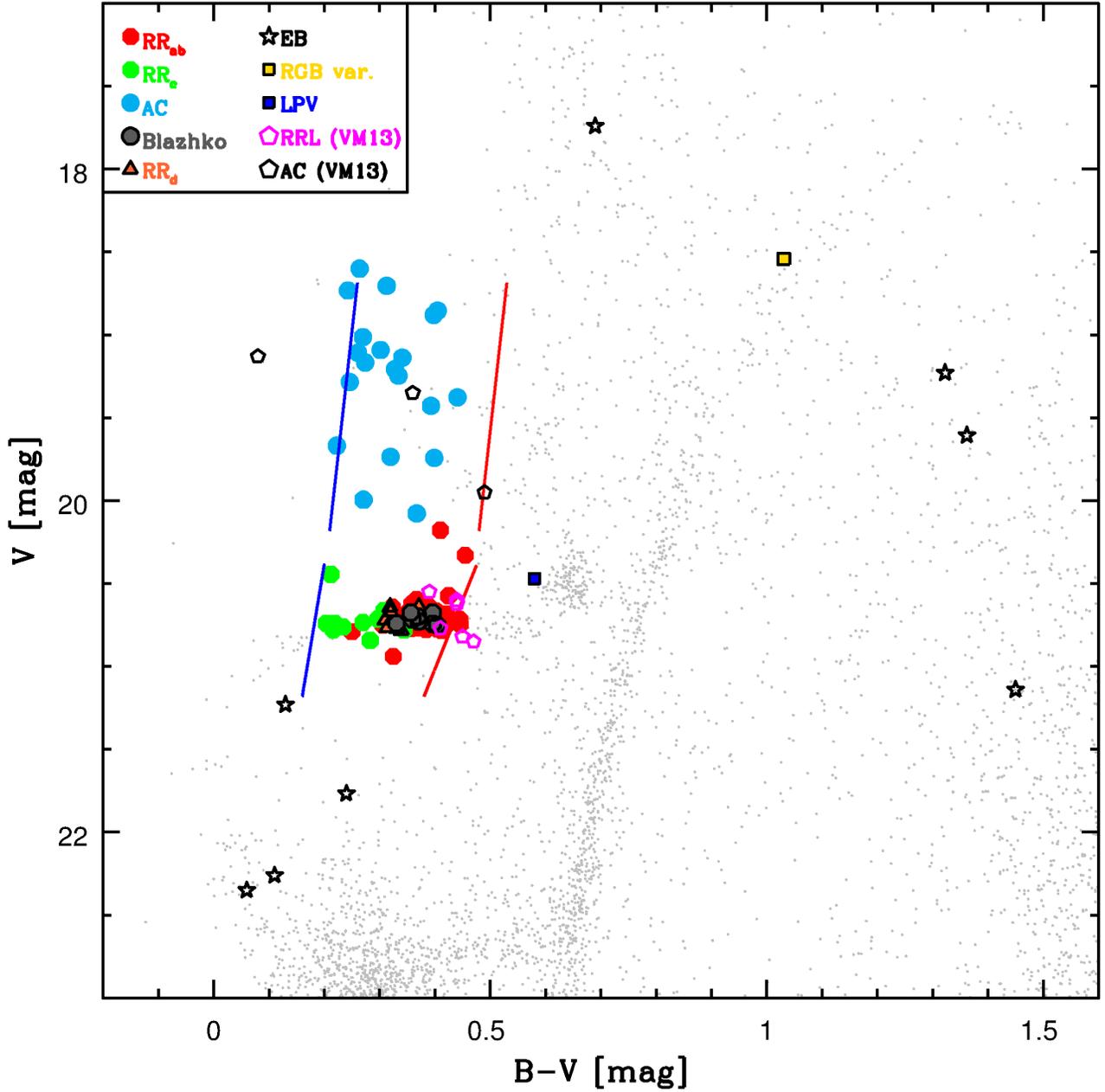}
\caption{Position of variable stars in $V$, $(B-V)$ CMD. Red and green circles display 
fundamental and first-overtone RRLs. 
Orange triangles and gray circles mark double-mode pulsators and candidate 
Blazhko RRLs. Cyan symbols show ACs. Yellow and blue squares display candidate RGB and LPV variables, 
while black stars show the EB variables. Two EBs, not members of
Carina, fall outside the plot limits; their colors and magnitudes are
given in Table~\ref{tab_prop}.}\label{fig:cmd} 
\end{figure*}

\begin{figure*}
\includegraphics[scale=0.9]{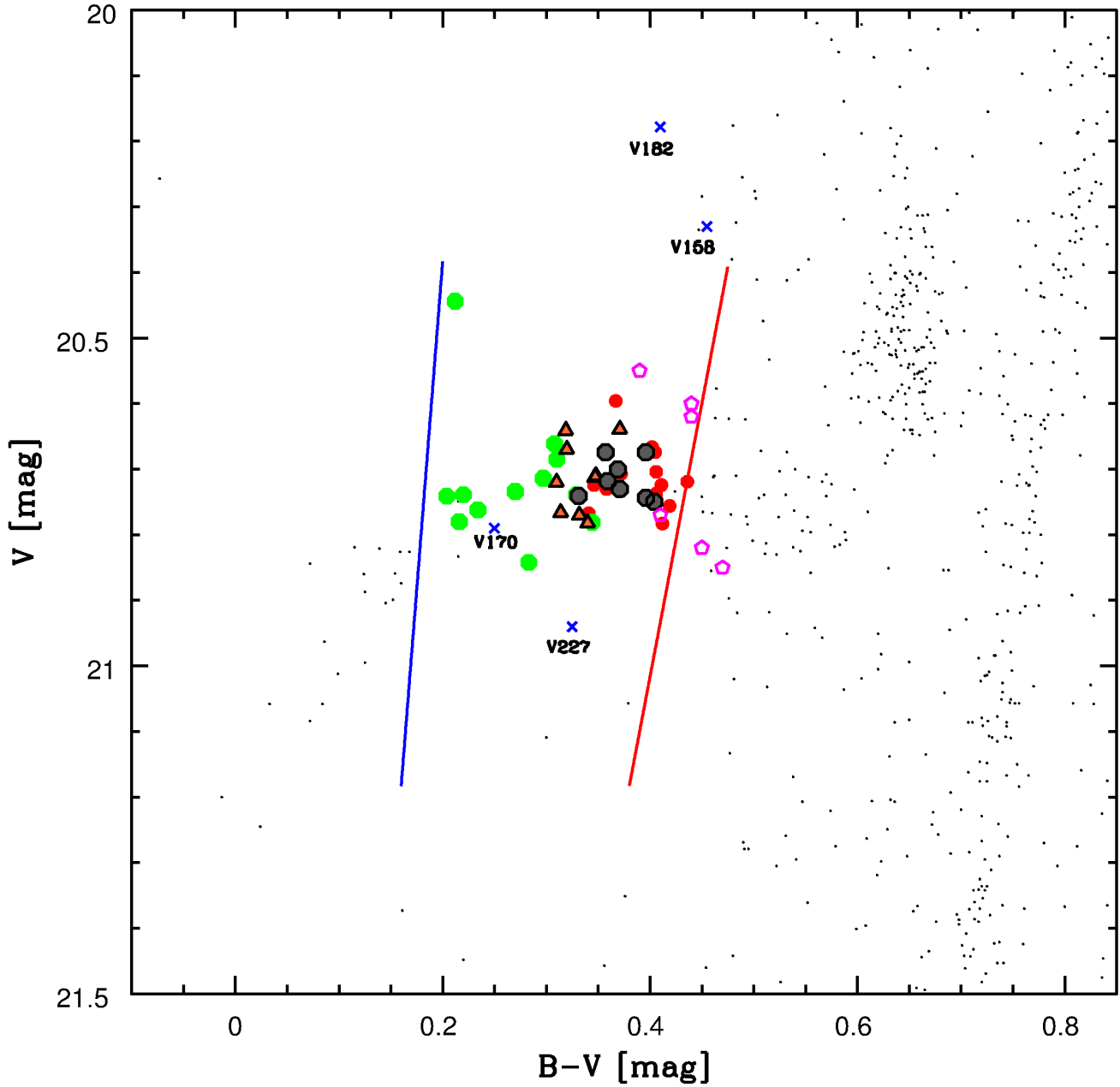}
\caption{Same as Figure~\ref{fig:cmd}, but zoomed on the CMD region located around the RR Lyrae 
instability strip.}\label{fig:cmd:bis} 
\end{figure*}

Figure~\ref{fig:cmd} and~\ref{fig:cmd:bis} show the instability strip
and a zoom of the position of RRLs in the $V$, 
$(B-V)$ CMD, respectively. Red and green circles display 
fundamental and first-overtone RRLs. Orange triangles and 
gray circles mark double-mode pulsators and candidate Blazhko RRLs, 
while cyan symbols show the location of AC variable stars. 
%The newly identified variables are represented by gray asterisks (Blazhko), 
%black crosses (RR$_{ab}$) and cyan crosses (AC). 
Yellow and blue squares represent RGB and LPV variables, while black stars 
display the locations of the EBs (see Table~\ref{tab_coordinates}). 
Finally, magenta and black pentagons show the RRLs and ACs in the
catalog of VMC13 and not in common with our catalog. Blue and red lines display the 
theoretical instability strip (IS) boundaries predicted by radial 
nonlinear convective pulsation models for ACs~\citep{fiorentino06} 
and RRLs~\citep[][hereinafter M2015]{marconi15}. 
Theoretical predictions were computed assuming a metal abundance 
of Z=0.0001 (ACs) and of Z=0.001 (RRLs).
Theory was transformed into the observational plane using the bolometric corrections 
and the color-temperature relations provided by~\citet{cardelli}. 
We also adopted a true distance modulus of  $\mu=(20.09\pm0.07)$ mag~\citep{coppola13} and a 
reddening of E({\it B--V\/})=0.03~\citep{monelli03}.  
The agreement between theory and observations appears, within the errors, 
quite good both for RRLs and ACs. 
In the above Figures the two bright RR$_{ab}$ stars (V158, V182) were defined 
as peculiar RRLs in Paper VI. We also note the presence of two RR$_{ab}$
(V170, V227) that appear anomalously blue. These stars have poorly sampled 
light curves and uncertain pulsational parameters. In particular, the number of 
measurements we have for the variable V170 is too small to fit the light curve 
and to estimate the luminosity amplitude (see Appendix for
details). These four stars are labelled and marked in Figure~\ref{fig:cmd:bis} with blue crosses. 

Moreover, we identified 101 new SX Phe. Their pulsation properties,
their light curves and their position in the $V$, 
$(B-V)$ CMD are discussed in \S~5.

%%%%%%%%%%%%%%%%%%%%%%%%%%%%%%%%%%%%%%%%%%%%%%%%%%%%%%%%%%%%%%%%%%%%%%%%%%%%%%%%%%%%%%%%%%%%
\section{Pulsation properties}

\begin{figure*}
\includegraphics[scale=0.6]{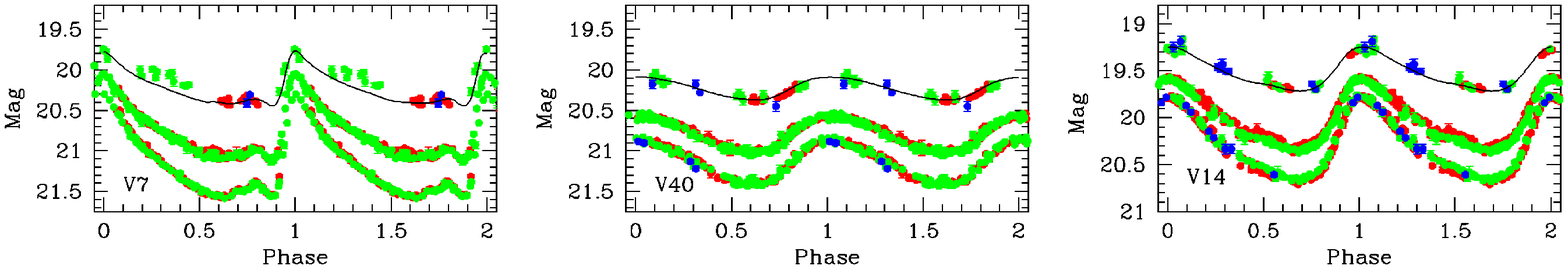} 
\caption{{\it BVI} light curves for selected Carina variables. We selected one
  RR$_{ab}$ (V7, left panel), one RR$_{c}$ (V40, middle panel) and one AC
  (V14, right panel). Red, green, blue and yellow filled circles display  
 different data sets: MOSAIC2$@$CTIO, WFI$@$MPI/ESO, Tek2K-I$@$CTIO and
  FORS1$@$ESO/VLT. Black lines are the fits of the light curves. For the
$I$-band they are the template light curves obtained by properly 
scaling the {\it V}-band light curve according to the procedure described in 
Section~\ref{sec:data}. The {\it BVI} plus the $U$-band  light curves 
for the the entire sample of variables are given in the electronic edition 
of the journal.}\label{fig:lc}
\end{figure*}

Figure~\ref{fig:lc} shows the {\it BVI} light curves for a fundamental (left) and first overtone 
(middle) RRLs plus an ACs (right). The different data sets were
plotted with different colors. 
The light curves of RR$_{d}$ variables are plotted in Figures~\ref{fig:rrd1}.
The complete atlas of light curves, including the $U$-band data, is available in the 
on–line edition of the paper. 

%_______________________________________________________________________________
\subsection{Bailey Diagram}\label{sec:bailey} 

\begin{figure*}
\includegraphics[scale=0.8]{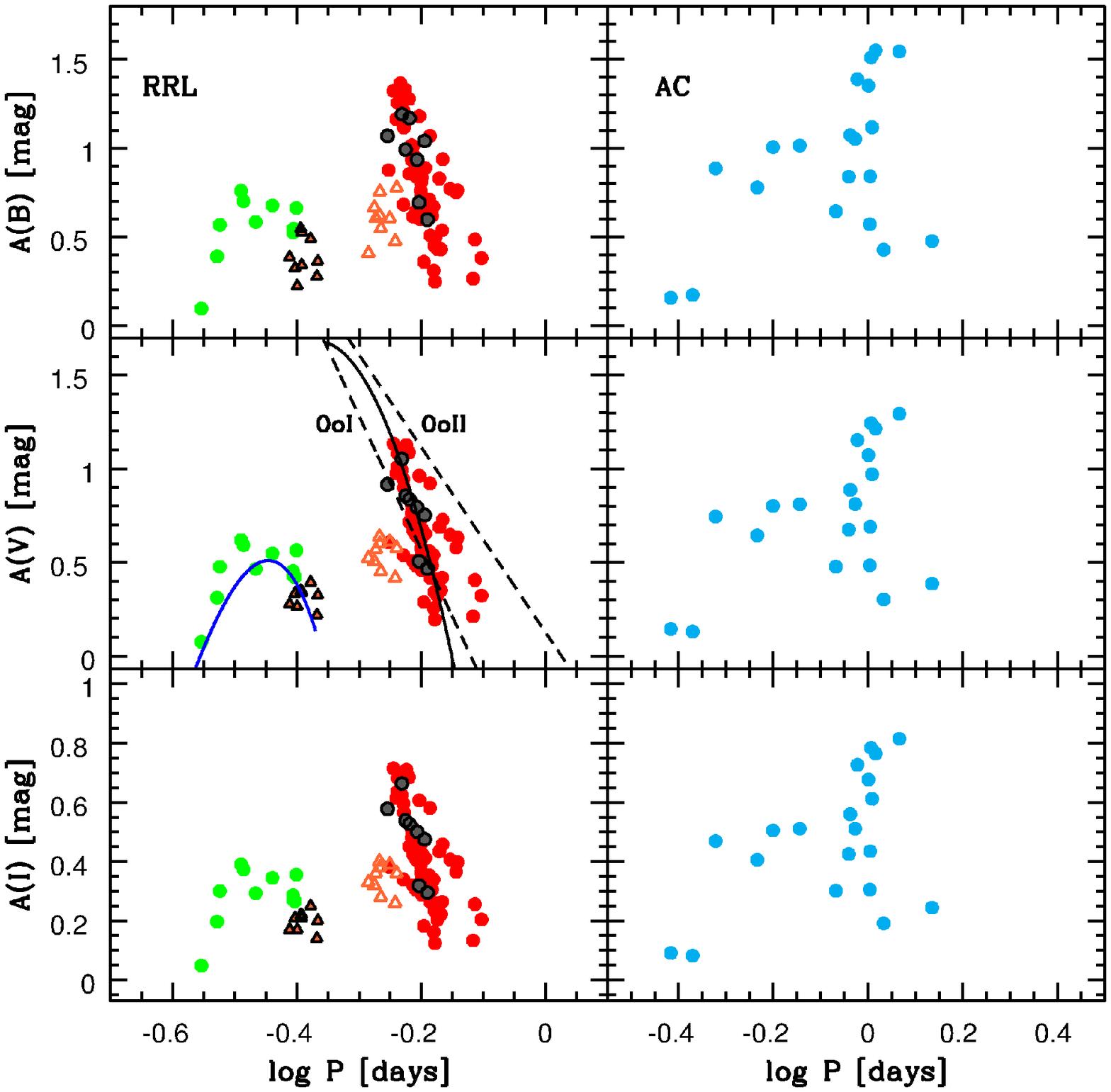} 
\caption{From top to bottom Bailey diagram in $B$- (top) $V$- (middle) and 
$I$-band (bottom) for Carina RRLs (left) and ACs (right). 
The dashed lines plotted in the middle left panel display OoI and OoII 
relations for fundamental pulsators in Galactic globular clusters 
according to~\citet{clement99} while the black solid line the relation for OoI 
cluster according to ~\citet{cacciari05}. The blue solid line shows the relation 
for OoII first overtone cluster variables provided by~\citet{kunder13}. 
Symbols are the same as in Figure~\ref{fig:cmd}. Note that double-mode variables 
have been plotted using periods and amplitudes of both primary (first
overtone, filled orange triangles) 
and secondary (fundamental, empty orange triangles) components.}\label{fig:bailey} 
\end{figure*}

The left panels of Figure~\ref{fig:bailey} display the Bailey diagram---amplitude vs 
period---for RRLs. From top to bottom the different panels show amplitudes in 
the $B$ (top), $V$ (middle) and $I$ (bottom) bands. The Bailey diagram is a 
very good diagnostic to split fundamental and first overtones. The transition 
period between low-amplitude, short-period RR$_{c}$ and large-amplitude, long-period 
RR$_{ab}$ is located around $\log P$$\sim$--0.35/-–0.30 ($P \sim 0.45-0.5$ days). The 
RR$_{ab}$ display a steady decrease in amplitude when moving toward longer 
periods. On the other hand, the RR$_{c}$ show a ``bell-shaped''~\citep{b97b}
or a ``hairpin''~\citep{kunder13} distribution. 
The Bailey diagram is also used to constrain the Oosterhoff
class~\citep[see, e.g.][]{b97a} of stellar systems hosting sizable samples 
of RRLs. The middle panel shows 
the comparison between RR$_{ab}$ in the $V$-band with the empirical relations 
provided by~\citet[][dashed lines]{clement99} and by~\citet[][black
solid line]{cacciari05}. The former relations are based on both OoI and OoII 
GCs, while the latter are only for OoI GCs. The comparison indicates that
Carina can be classified either as an OoI or as an Oosterhoff intermediate 
system, since for periods longer than $\log P$$>$-0.2 there is a group of RR$_{ab}$ 
that, at fixed period, attains larger luminosity amplitudes. On the other hand, 
the amplitudes typical of RR$_{c}$ variables in OoII GCs~\citep[][blue
solid line]{kunder13} agree quite well the current data. 

In Paper VI we classified Carina, on the basis of the mean period of 
fundamentalized RRLs, as an OoII system. Indeed, we found  
$<P_{RR}>=0.60\pm0.01$  days ($\sigma$=0.07). 
Including the new discovered RR$_{ab}$ variables we find that the mean 
period of fundamental RRLs is $<P_{RR}>=0.637\pm0.006$  days ($\sigma$=0.05). 
This estimate suggests that Carina is closer to an OoII stellar systems, since 
OoII GCs show mean fundamental periods of $< P >\,\sim\, 0.65 \,$days, while 
the OoI GCs show shorter periods, namely $< P >\,\sim \,0.55 \,$days. 
We have also computed the ratio between the number of RR$_{c}$ and the 
total number of RRLs, and we found $N_{c}/(N_{ab}+N_{c}) \sim 0.14$, 
i.e., a fraction of RR$_{c}$ variables that is more typical of OoI 
($\sim\,$17\%) than OoII ($\sim\,$44\%) GCs. 
The above evidence indicates that the Oosterhoff classification of Carina 
depends  on the adopted diagnostic. This means that the Oosterhoff 
classification should be cautiously treated, since it depends on the 
adopted diagnostic on the completeness and size of the RRL sample and on the morphology of the Horizontal Branch~\citep{fiorentino15a}. 

\begin{figure*}
\includegraphics[scale=0.9]{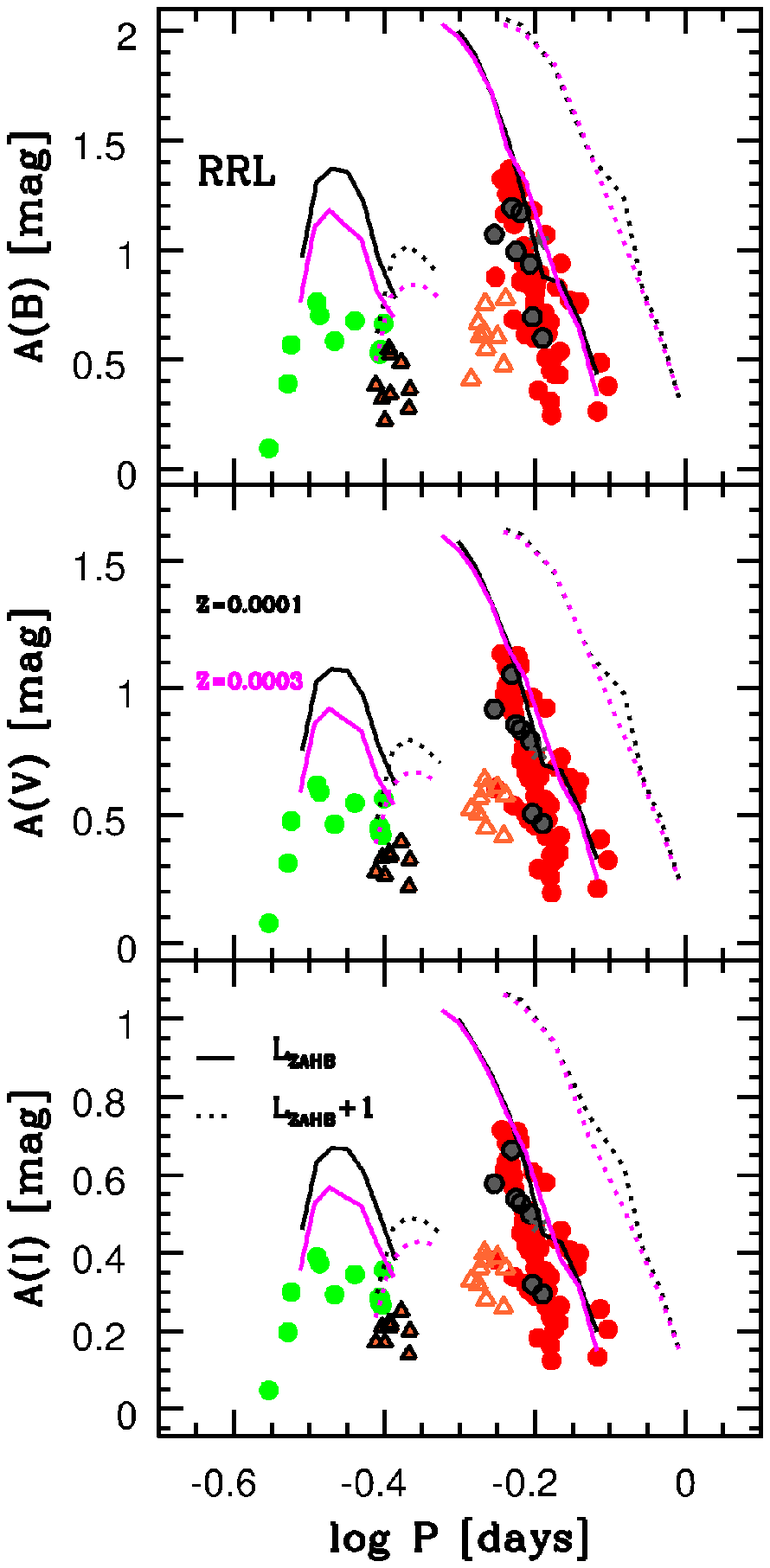} 
\caption{Same as left panels of Figure~\ref{fig:bailey}, but the
  comparison is between observations and predicted luminosity
  amplitude provided by M2015. The solid lines display predicted
  amplitude for F and FO models constructed by assuming a stellar mass
  of M=0.80 M$_{\odot}$ a metal poor chemical composition (Z=0.0001,
  Y=0.245) and Zero-Age-Horizontal-Branch luminosity level ($\log (L\L_{\odot})$=1.76). 
The black dotted lines display the same predictions, but for models
constructed by assuming a brighter luminosity level  ($\log
(L/L_{\odot}$)=1.86). The purple lines display the same pulsation
predictions, but for pulsation models constructed by assuming a less
metal-poor chemical composition (Z=0.0003, Y=0.245).}\label{bailey_model}
\end{figure*}

The data plotted in Figure~\ref{fig:bailey} also show mixed-mode pulsators, 
the so-called RR$_{d}$ variables.
These are radial variables oscillating simultaneously in at least two 
different pulsation modes. The RR$_{d}$ typically oscillate in the 
first overtone and in the fundamental mode, and the former mode 
is usually stronger than the latter~\citep[e.g.][]{smith2006}, 
but there are exceptions~\citep{clementini04}. 
On the other hand, Classical Cepheids display a wide range of mixed-mode 
pulsators among the overtones and fundamental mode~\citep{soszynski08,soszynski10}, 
suggesting that surface gravity and effective temperature might play fundamental       
roles in driving the occurrence of such a phenomenon. 
Figure~\ref{bailey_model} shows the comparison in the Bailey diagram
between the current observations and predicted amplitudes. Pulsation
prescriptions rely on a large set of RRL models recently provided by
M2015. The black solid and dotted lines display predicted amplitudes for
the sequence of metal-poor (Z=0.0001, Y=0.245) models constructed by
assuming a stellar mass of 0.80 M$_\odot$ and two different luminosity
levels. The black solid line shows predictions for the
Zero-Age-Horizontal-Branch (ZAHB) luminosity level ($\log
(L/L_{\odot})$=1.76), 
while the dotted line for a brighter luminosity level ($\log
(L/L_{\odot}$)=1.86). 
The purple lines display the same predictions, but for
a slightly more metal-rich chemical composition
(Z=0.0003, Y=0.245; M=0.716 M$_{\odot}$, $\log (L/L_{\odot}$)=1.72 and 1.82 ).

The predicted amplitudes appear to be slightly larger, at fixed
pulsation period, when compared with observations. This is a limit in
the current theoretical framework, since the amplitudes are tightly
correlated with the efficiency of convective transport. In passing we
note that the comparison between predicted and observed 
luminosity amplitudes is hampered by the current theoretical uncertainties 
in the treatment of the time dependent convective transport. We current adopt 
a mixing length parameter of $\alpha$=1.5~\citep[see][]{marconi11}. Larger values cause a steady decrease in the 
luminosity amplitude~\citep[see][]{dicriscienzo04}. Indeed, observed amplitudes attain, at
fixed period smaller amplitudes. This applies to both fundamental and
first overtone variables.  Moreover, we are assuming that 
observed light curves have a good sampling around the phases of minimum and 
maximum light. However, the comparison shows two
interesting features. i) The predicted amplitudes for FO pulsators
shows a larger dependence on metal content than fundamental
pulsators. ii) The regular pulsators display, at fixed pulsation
period, a small spread in amplitudes. The current predictions suggest
that their evolutionary status is quite homogenous, since they appear
to be located to the ZAHB luminosity level.

Carina was previously known to host six RR$_{d}$ variables: V11, V26,
V89, V192, V198, V207 (D03), in the current analysis we discovered
other three double-mode pulsators: V74, V210 and V225.
To further understand their nature, and in particular to properly define
the location of RR$_{d}$ pulsators in the Bailey diagram, we estimated
both primary (first overtone) and secondary (fundamental) periods and
decomposed their light curves. Indeed, the quality of the photometry
allowed us to estimate not only the ``global luminosity amplitude'',
but also the amplitude of both fundamental (open orange triangles) and
first overtone mode (filled orange triangles).
Table~\ref{tab_prop_rrd} gives from left to right their periods,
mean magnitudes and amplitudes in the {\it BVI} bands.
To our knowledge this is the first time in which we can associate
to the two modes of double-mode variables their individual luminosity
amplitudes.
The data plotted in the left panels of Figure~\ref{fig:bailey} bring forward even to a
cursory scrutiny that the primary components (first overtone) are located
in the long period tale ($\log P\sim$-0.4) of single mode first overtone
variables. Moreover, the secondary components (fundamental) are located
in the short period tale ($\log P\sim$-0.25) of single mode fundamental
variables.
This evidence is further 
supported by their mean colors. The mean color of RR$_{d}$ is 
systematically redder ($B-V$$\sim$0.33 mag) than the color range covered by 
RR$_{c}$ variables (0.2 $\le$ $B-V$ $\le$ 0.35 mag) and systematically bluer  than the 
typical color range of RR$_{ab}$ variables  (0.3 $\le$ $B-V$ $\le$ 0.5 mag).

The pulsation and evolutionary status of the RR$_{d}$ variables 
depends on their evolutionary direction and on their position in 
the so-called OR region~\citep{b97a}. In this context it is worth mentioning that 
RRLs in dwarf spheroidal galaxies appear to lack High Amplitude Short Period (HASP) 
fundamental variables~\citep{stetson14a, fiorentino15a}, thus resembling
Oosterhoff II globular clusters in the Milky Way~\citep{b97a}.
 
The occurrence of a good sample of RR$_{d}$ variables in Carina seems to suggest 
that this region of the Bailey diagram might be populated by mixed-mode 
variables. Obviously, the occurrence of RR$_{d}$ variables depends on the 
topology of the instability strip, but also on the evolutionary 
properties (extent in temperature of the so-called blue hook) and, in 
particular, on the occurrence of the hysteresis
mechanism~\citep{vanalb71,bono95,fiorentino15a,marconi15} when moving 
from more metal-poor to more metal-rich stellar structures.     

The Blazhko RRLs plotted in Figure~\ref{fig:bailey} also appear to be located  
in a very narrow period range. No firm conclusion can be reached concerning 
the distribution in the Bailey diagram of Blazhko RRLs, since the sample size 
is quite limited and also because the Blazhko cycle is poorly sampled. 

The above findings further support the crucial role played by cluster and 
galactic RRLs to constrain the topology of the instability strip and to investigate 
the evolutionary and pulsation status of exotic objects like RR$_{d}$ and Blazhko RRLs.       

The right panels of Figure~\ref{fig:bailey} show the distribution of Carina ACs in the 
same Bailey diagrams as the RRLs. Their properties have already been discussed in 
Paper VI. We confirm the separation at $\log P$$\sim$-0.1 between long-period and 
high-amplitude with short-period and low-amplitude ACs. In passing we note that 
three out of the 20 ACs have periods around one day. Thus 
further supporting the need for data sets covering large time intervals to remove 
the one-day alias. 

Note that in the current analysis of evolved variables we did not include
the six RRLs (three RR$_c$, three RR$_{ab}$) and the three ACs
recently detected by VM13
outside the tidal radius of Carina. The reasons are
the following. The three RR$_c$ variables attain periods that are systematically
shorter (-1.0 $\lesssim \log P \lesssim$ -0.25) that typical RR$_c$ variables
(see Table~\ref{tab_prop_vivas}), but their mean $B-V$ colors are typical of RR$_{ab}$ variables.
The same outcome applies to the three RR$_{ab}$ variable, and indeed they are located
in the short period range of fundamental pulsators, but their $B-V$ colors are
typical of objects located close to the red edge of the instability strip.
Moreover, the newly identified ACs have periods that are
systematically shorter (-0.8 $\lesssim \log P \lesssim$ -0.7) than the typical
Carina ACs.

%_______________________________________________________________________________
\subsection{Petersen diagram}\label{sec:rrd}

\begin{figure*}
\includegraphics[scale=0.9]{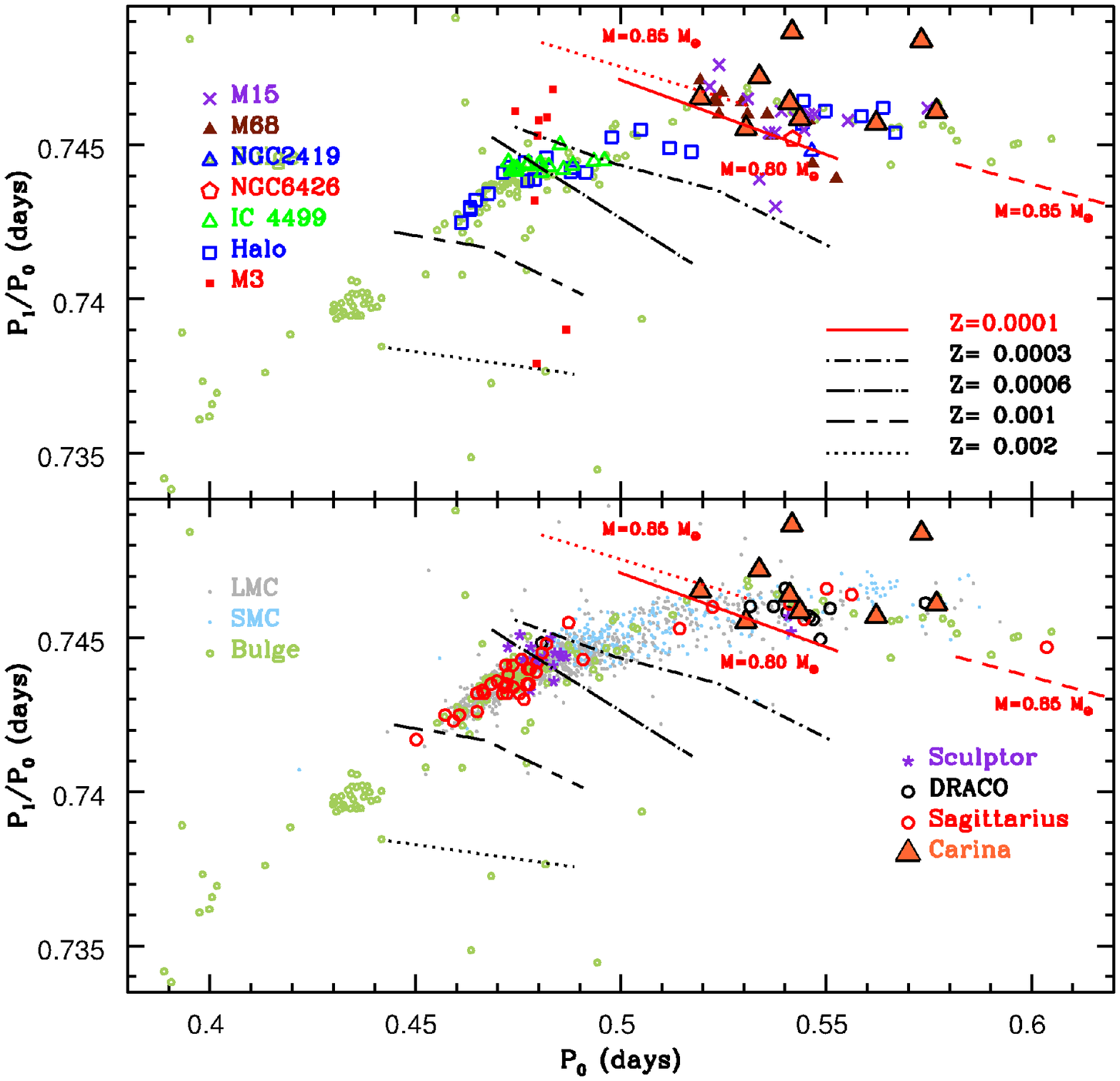} 
\caption{Top -- Comparison in the Petersen diagram between Galactic 
(halo, bulge, globular clusters) and Carina (orange triangles) RR$_{d}$ 
variables.  Pulsation predictions (M2015) for different chemical compositions 
(see labeled values) are plotted with different lines.
The stellar mass of the pulsation models for the most metal-poor chemical 
composition are also labeled.  
Bottom -- same as the top, but the comparison is with RR$_{d}$ variables 
in nearby dwarf spheroidals and irregulars. The RR$_{d}$ of the Galactic 
bulge are also plotted.}\label{petersen2}
\end{figure*}

The top panel of Figure~\ref{petersen2} displays the position of the Carina 
$RR_d$ variables in the Petersen diagram, i.e., the first-overtone-to-fundamental 
period ratio ($P_1/P_0$) versus the fundamental period. The data in this panel 
also show the comparison with $RR_d$ variables identified in Galactic globulars 
(see labeled names) and in the Galactic field (Halo: blue squares; 
Bulge: small green circles). Table~\ref{tab_rrd_other} gives from left to right the name of 
the stellar system, the number of $RR_d$ variables, the reference for the 
$RR_d$ data, the mean metallicity and the reference for the metallicity 
estimate.    
The data plotted in this figure display several interesting features.

The period ratio shows a steady decrease when moving 
from more metal-poor to more metal-rich systems~\citep[see also][]{bragaglia01}. 
The largest values in the period ratio are attained in the Halo and in very 
metal--poor GCs, while the smallest in the Bulge. The trend with the metallicity 
was also suggested on both theoretical and empirical bases 
by~\citet{soszynski11} and more recently by~\citet{soszynski14}. Note that 
the fraction of double--mode variables appears to be anti-correlated with 
the mean metal abundance, and indeed it increases from 0.5\% in the Bulge 
to 4\% in the LMC and to 10\% in the SMC~\citep{soszynski11}. The RR$_d$ 
variables in Carina appear to follow a similar trend, since the current 
fraction is $\sim$10\%. Two RR$_{d}$ variables (V192, V210) display
period ratios that are systematically 
larger then period ratios in Galactic globulars, in dwarf galaxies and in the 
Bulge. In passing we note that one RR$_d$ in M68 (V36) and two in M15 (V51, V53) 
display period ratios that systematically smaller than the 
bulk of RR$_d$ variables. The referee suggested to constrain on more quantitative 
basis the difference. To avoid spurious fluctuations in the period range 
covered by the different data sets, we ranked the entire sample (398) as a 
function of the fundamental period. Then we estimated the running average 
by using a box including the first 40 objects in the list. We estimated the 
mean period ratio, the mean fundamental period and the standard deviations of 
this sub-sample. We estimated the same quantities by moving of one object in 
the ranked list until we took account of the last 40 objects in the sample. 
We performed several tests changing both the number of objects included in
the box and the number of stepping stars. The current finding are minimally 
affected by plausible variations. We found that the quoted five RR$_d$stars are 
located within 3$\sigma$ of the mean. The current statistics is too limited 
to claim solid evidence of discrepancy. Note that we double checked the 
photometric quality and coverage of the above five objects and we found that 
they are similar to the other canonical RR$_d$stars.  

The period ratios display a larger spread when moving into 
the metal-poor regime (long fundamental periods). To define the trend
on a more quantitative basis we adopted several sets of nonlinear, 
convective pulsation models (M2015). 
The adopted chemical compositions are labeled.
The new set of RRL models 
relies on the theoretical framework outlined
in~\citet{dicriscienzo04} and~\citet{marconi11}, but on new evolutionary 
prescriptions for low-mass He burning models provided by~\citep{pietrinferni04}
(BASTI data base, http://albione.oa-teramo.inaf.it). The pulsation predictions plotted in this 
panel cover the so-called OR region, i.e., the region in which RRLs show 
a stable limit cycle in both the fundamental and the first overtone mode.     
This region is located between the blue edge of the fundamental mode and the red 
edge of the first overtone. 
The comparison between theory and observations indicates that an increase 
in the luminosity ($\log L/L_\odot$=1.76, 1.86, dotted and dashed red lines) 
mainly causes, at fixed chemical composition (Z=0.0001) and stellar mass 
($M/M_\odot$=0.85), a steady increase in the fundamental period, 
and in turn a decrease in the period ratio. Moreover, a decrease in stellar 
mass ($M/M_\odot$=0.80, solid red line) at fixed chemical composition and 
luminosity level ($\log L/L_\odot$=1.76) causes a systematic decrease in the period ratio 
and a moderate increase of the fundamental period. 
The above trends take account of a significant fraction of RR$_{d}$ pulsators 
located in metal-poor GCs and in Carina dSph. This indicates that RR$_{d}$ in 
Carina have a metallicity of the order of Z=0.0001 and a mean stellar mass 
close to $0.85 M_{\odot}$. 
The pulsation masses are slightly larger than predicted by evolutionary 
models, but are within the current empirical and theoretical
uncertainties. 

%Note that the adopted luminosity levels do not take account of the new 
%conductive opacities suggested by~\citet{cassisi07}\footnote{The
% effect of using the new opacities for HB calculation is to decrease the bolometric luminosity
% of the ZAHB by $\log(L/L_{\odot}) \sim 0.002$.}. 

The above findings further support the evidence that 
the old stellar component in Carina is quite metal-poor. This 
evidence is soundly supported by recent photometric and spectroscopic 
results by~\citet{monelli14} and~\citet{fabrizio15} suggesting a mean 
metal abundance for the old stellar component of [Fe/H]=(--2.13$\pm$0.03$\pm$0.28) dex.  

In this context it is worth mentioning that theoretical predictions 
for more metal-rich pulsation models (see black lines and labeled values) 
provide a sound explanation of the steady decrease in the period ratio of 
Bulge RR$_{d}$ variables, i.e., the stellar systems with the broader metallicity 
distribution. 

To further define the pulsation and evolutionary properties of Carina 
RR$_{d}$ variables, the bottom panel of Figure~\ref{petersen2} shows the same 
Petersen diagram, but the comparison is now extended to RR$_{d}$ in nearby 
dwarf spheroidals (Draco, Sculptor, Sagittarius), in dwarf irregulars   
(Large Magellanic Cloud, LMC; Small Magellanic Cloud, SMC) and in the 
Bulge. The data plotted in this panel bring forward several interesting 
new findings. 

The range in period ratios covered by LMC RR$_{d}$ is on 
average larger (0.740 $\le$$P_1/P_0$$\le$ 0.749) than the 
range of SMC RR$_{d}$ variables. This evidence supports spectroscopic 
measurements of LMC RRLs suggesting metal abundances ranging from [Fe/H]=
--2.12 to --0.27 dex~\citep{gratton04}. The SMC RR$_{d}$ cover a slightly 
narrower period ratio range (0.741 $\le$$P_1/P_0$$\le$ 0.747 days) but
according to recent studies, and within current uncertainties
affecting metallicity estimates for RRLs in these systems~\citep{h2012}
the metallicity spread for the old stellar populations
in the two Clouds is similar. 

The location of RR$_{d}$ of Carina and Draco is the same in the
Petersen diagram. Indeed current spectroscopic estimates, based on medium 
resolution spectra, provide a very metal-poor iron abundance
([Fe/H]=--1.92$\pm$0.01 dex, see Table~\ref{tab_rrd_other}) also for Draco. 
The steady decrease in period ratio of RR$_{d}$ in Sculptor 
is strongly supported by the recent spectroscopic measurements suggesting 
an iron abundance of [Fe/H]=--1.68$\pm$0.01 dex (see Table~\ref{tab_rrd_other}).      
The empirical evidence concerning Sagittarius needs to be discussed in detail, 
because the periods and period ratios attain values that are on average smaller 
than for RR$_{d}$ in other dSphs. Spectroscopic estimates based on 
high-resolution spectra by~\citet{carretta10}, suggest for Sagittarius 
a mean iron abundance, based on 27 RGs, of [Fe/H]=--0.62 
and individual values ranging from -1.0 to above solar. A smaller spread in iron 
abundance was also suggested by \citep[][]{kc08} using RR Lyrae properties. 
The spread in period ratios and the range in fundamental periods 
(0.45 $\le$P$\le$ 0.49 days) showed by Sagittarius RRLs soundly support the 
spectroscopic measurements and the similarity with LMC RRLs. This indicates that 
Sagittarius is a fundamental nearby laboratory to constrain the pulsation properties 
of metal-rich RRL in gas poor systems. 

The RR$_{d}$ in the Bulge (small green circles) display a clear overdensity 
for $P_0$$\sim$0.46 days. This overdensity was explained by~\citet{soszynski11} 
as the relic of a former dwarf galaxy that was captured by the Milky Way. 
More recent investigations based on a larger sample (28 vs 16) indicates 
that they are distributed along a stream that crosses 
the Galactic bulge almost vertically. Note that the comparison with theoretical predictions  
suggests, for the above stellar system, a metal-intermediate chemical 
composition (Z=0.001--0.002). 

The RR$_{d}$ also provide a unique opportunity to validate the current approach 
to fundamentalizing the first overtones. Whenever the sample of RRLs hosted 
in a stellar system is limited, fundamental and first overtone variables 
are treated as a single sample by transforming the periods of the first overtones 
into "equivalent" fundamental periods,
using the relation $\log P_F$=$\log P_{FO}$ + 0.127. This assumption dates 
back to almost half a century and relies on the few RR$_{d}$ variables known 
at that time~\citep{sandage81, cox83, petersen91}. The above constant period shift is 
further supported by the new theoretical scenario by M2015 and by the sizable 
sample of RR$_{d}$ variables recently identified by large, dedicated photometric 
surveys. 

\begin{figure*}
\includegraphics[scale=0.9]{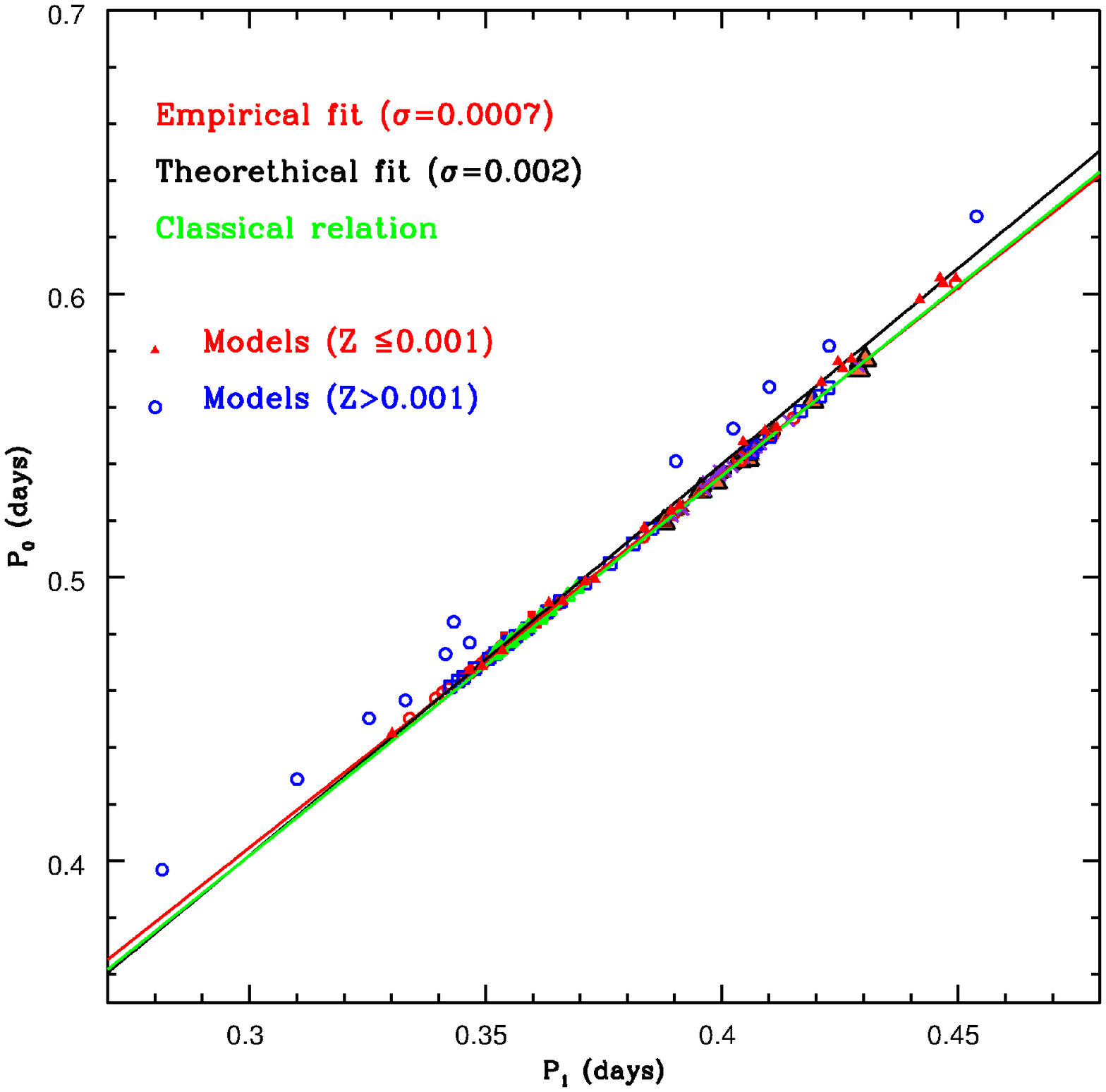} 
\caption{Correlation between fundamental and first overtone
periods. Observed RR$_d$ variables have been plotted using the same
symbols of Figure~\ref{petersen2}. The red line shows the fit to the 
observed data, the black line the fit to the pulsation models and the 
green line the classical relation. Red triangles and blue dots display 
metal--poor and metal--rich pulsation models.}\label{petersen}
\end{figure*}

In passing we note that this issue is far from being an academic  
dispute, since the same fix is also used to improve the precision 
of distance determinations based on both RRLs~\citep{braga15} 
and classical Cepheids~\citep[e.g.][]{marengo10}. 
Figure~\ref{petersen} shows fundamental vs first overtone periods ($P_0$ vs $P_1$) 
for the entire sample of RR$_{d}$ plotted in Figure~\ref{petersen2}. The red line shows 
the fit to the empirical data. We found:
\begin{equation}\label{eq:1}
P_0 = 0.0096+1.317*P_1 ;
\end{equation}

The new estimate of both the slope and the zero-point soundly supports the 
old fix, and indeed the green line, showing the classical fix, agrees 
quite well with the new observations. 

To further constrain the impact that the period ratios of RR$_{d}$ variables 
have in constraining the pulsation properties of RRLs, we plotted in the 
same plane theoretical predictions for the "OR" regions adopted in Figure~\ref{petersen2}.  
The black solid line was estimated by considering only models 
more metal-poor than Z=0.001, i.e., a metallicity range similar to the 
observed one. The agreement is quite good over the period range. 
In passing we also note that when moving to more metal-rich models (Z$>$0.001, blue 
circles) the period increase is significantly larger in the
fundamental period than in the first overtone one. We still lack firm empirical 
evidence for such objects and it is not clear whether it is an 
observational bias or the consequence of an evolutionary property
connected to the dependence of the HB morphology on the metal content.

%____________________________________________________________________________________
\section{Distance to Carina from optical Period--Wesenheit relations}
\subsection{Carina distance determination based on the empirical  PW
  {\it BV\/} relation }
\begin{figure*}
\includegraphics[scale=0.85]{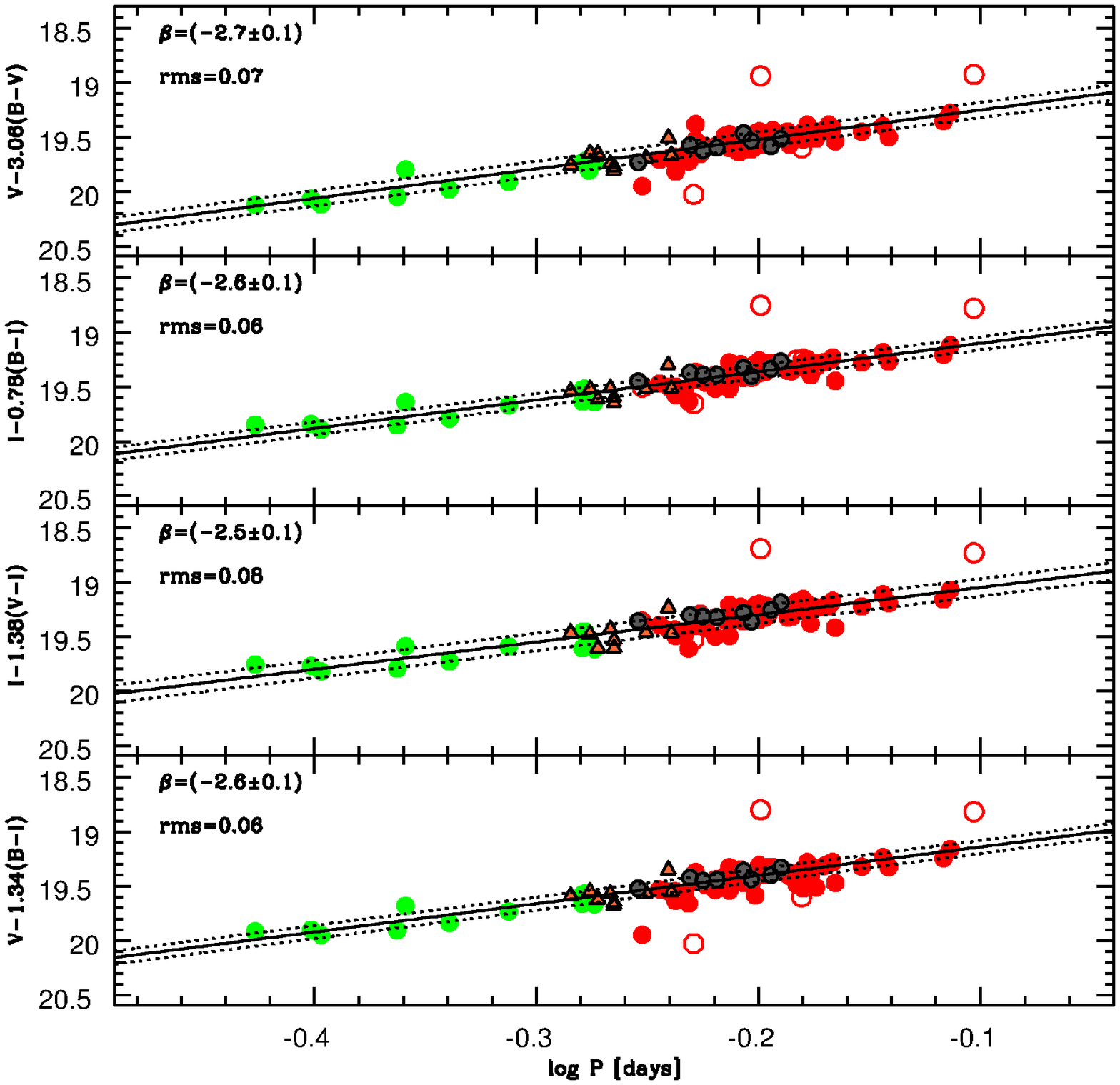} 
\caption{Observed optical PW relations. From top to bottom the different panels 
display the {\it BV}, the  {\it BI}, the  {\it VI} and the {\it BVI} relations. First overtone pulsators were fundamentalized. The symbols 
are the same as in Fig.~\ref{fig:cmd}. The solid lines display the fit, while the dotted lines 
the 1$\sigma$ difference. The standard deviations (rms), the coefficient of
the logarithmic period and their errors are also labeled. The two empty 
circles were not included in the estimate of the PW relations.}\label{fig:plw2} 
\end{figure*}

One of the most important tools for deriving distances from
pulsating stars is the so called Wesenheit relation~\citep[see for
example][]{vdb75,madore82} that is independent of reddening by
definition, assuming that the ratio of total-to-selective
absorption is fixed. This is a period-luminosity relation that
includes a color term whose coefficient is the ratio between the total
and the selective extinction coefficients. 
Figure~\ref{fig:plw2} shows the observed PW
relations and the empirical fit to the data (solid black lines)
obtained by fundamentalizing the first overtone pulsators by using Equation~\ref{eq:1}. Dashed lines
depict the dispersion of the above inferred relations. Results of
these fits are listed in Table~\ref{table:wesenheit1}, where the
zero-points, the slopes and the dispersions of the relations are reported in the
first three columns, respectively.  In the fit determination we excluded stars outside 3$\sigma$ of
the inferred empirical {\it BV\/} Wesenheit relation. These stars are the
two peculiar pulsators V158 and V182 and the stars
V170 and V171 for which we have uncertain parameters (red open symbols).

\begin{figure*}
\includegraphics[scale=0.6]{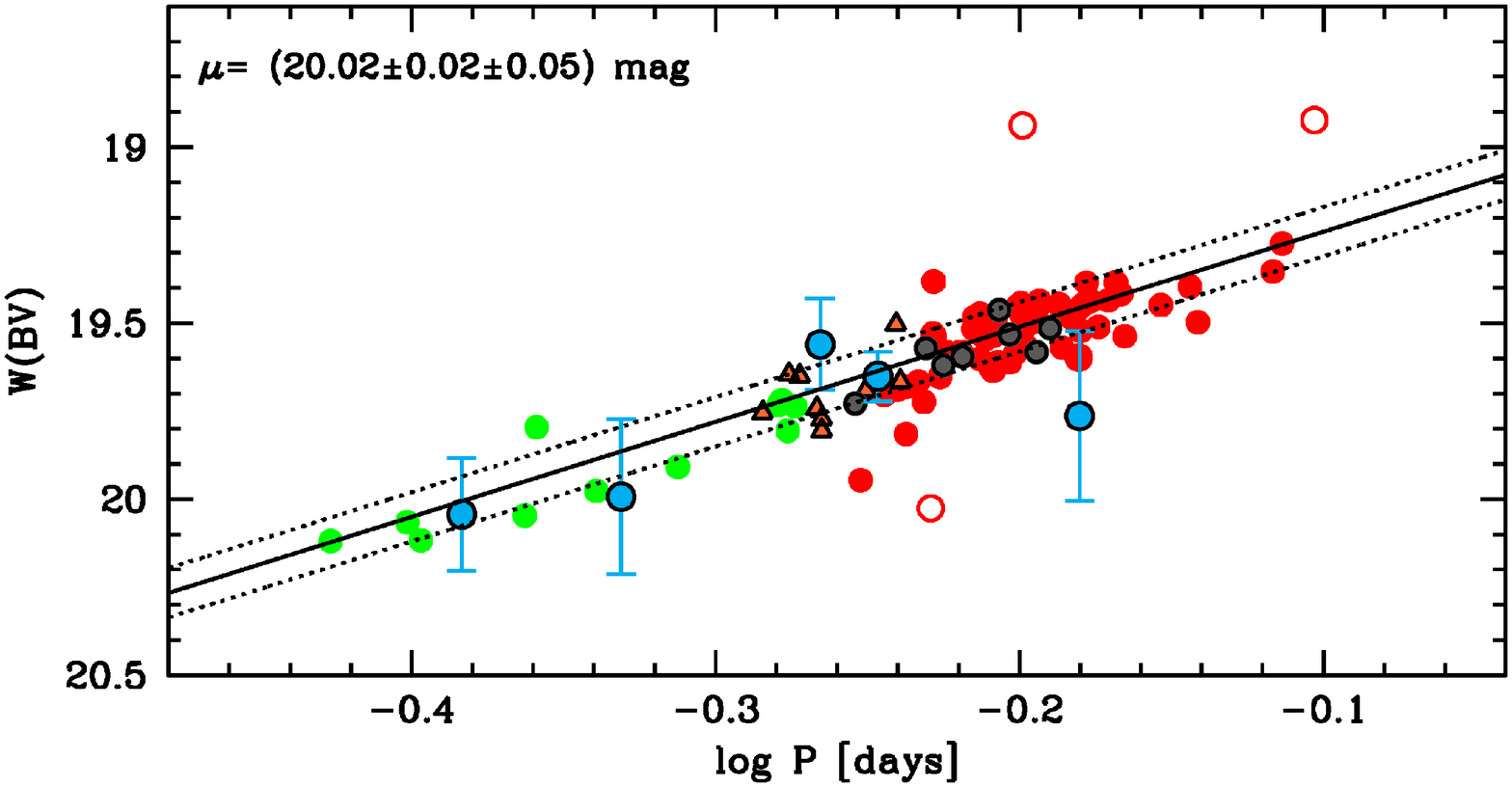} 
\caption{Same as the top panel of Figure~\ref{fig:plw2}, but with the the five 
field RRLs (light blue circles) for which trigonometric parallaxes have been 
estimated using FGS at HST. The vertical error bars take account of the 
photometric error and of the uncertainty in distance. The object with the 
smallest error bar is RR Lyr itself. The true distance modulus based on 
the empirical slope and on the calibrating RRL is labeled.}\label{fig:calibrating} 
\end{figure*}

Thanks to the use of the Fine Guidance Sensor on board the HST,~\citet{benedict11}
provided accurate estimates of the
trigonometric parallaxes for five field RRLs: SU Dra, XZ Cyg, RZ Cep,
XZ Cyg and RR Lyr.
Using their data in Table~2, we derived the mean magnitude in the $B$ and $V$ bands from a fit
with a spline under tension. We then calculated the
absolute Wesenheit parameter for each star, ($W=<V>-3.06(B-V) -
\mu$), where $\mu$  is the individual distance modulus based
on the HST parallax. Individual mean magnitudes of the
calibrating RRLs and their distances are listed in Table~\ref{tab_calibrating}. We applied
the individual calibrating RRL to the empirical PW relations (see
Figure~\ref{fig:plw2} and Table~\ref{table:wesenheit1}). Note that the
calibrating RRLs cover a
limited range in metallicity~\citep[from -1.80 to -1.41 dex,][]{benedict11}. The current theoretical
predictions (see Subsection~\ref{sub:teo_dist}) suggest a mild dependence on the
metal content. Therefore, we neglected the metallicity dependence of
the calibrating RRLs.
The data plotted in Figure~\ref{fig:calibrating} show in the W-log P plane the calibrating RRLs
together with the Carina RRLs. The error bars of the calibrating RRLs
take into account both the photometric errors and the uncertainties of
the trigonometric parallaxes. A glance at the data discloses that only
for RR Lyr is the precision of the absolute distance better than
1~$\%$. Using all the calibrating RRLs we found a true distance modulus
for Carina of $\mu= (20.02 \pm 0.02 \pm 0.05)$ mag. Note that for the above reasons the accuracy of
the distance mainly depends on the accuracy of the data for RR Lyr.

\subsection{Carina distance determination based on theoretical PWZ
  relations }\label{sub:teo_dist}

To fully exploit the multiband photometry of Carina RRLs we also
decided to use predicted PW relations. The new theoretical framework
derived by M2015 indicates that the PW
({\it BV}) relation is independent of the metal content. This is a very
positive feature for two different reasons: a) the PW ({\it BV}) can be applied
to estimate individual distances of field RRLs for which the metal
content is not available; b) the PW ({\it BV}) is particular useful to
estimate distances of RRLs in  nearby dwarf galaxies, since they
typically cover a broad range in iron content. 
We applied the predicted relation to Carina RRLs and we found a true
distance modulus of $\mu = (20.08 \pm 0.007 \pm 0.07)$ mag. The distance determination has been estimated
using the predicted relation for fundamental pulsators. The number of
RR$_{c}$ variables in Carina is modest (12) and they have been
fundamentalized.
As noted in the previous section in the current distance
determination we did not consider stars outside 3$\sigma$ of
the inferred empirical Wesenheit  ({\it BV\/}) relation. The above
distance agrees, within the errors, quite well with the true Carina
distance based on empirical calibrators. Moreover, the new distance
determination also agrees with Carina distances available in the
literature that are based on robust standard candles (see Table~\ref{tab_car_distance}).

\begin{figure*}
\includegraphics[scale=0.85]{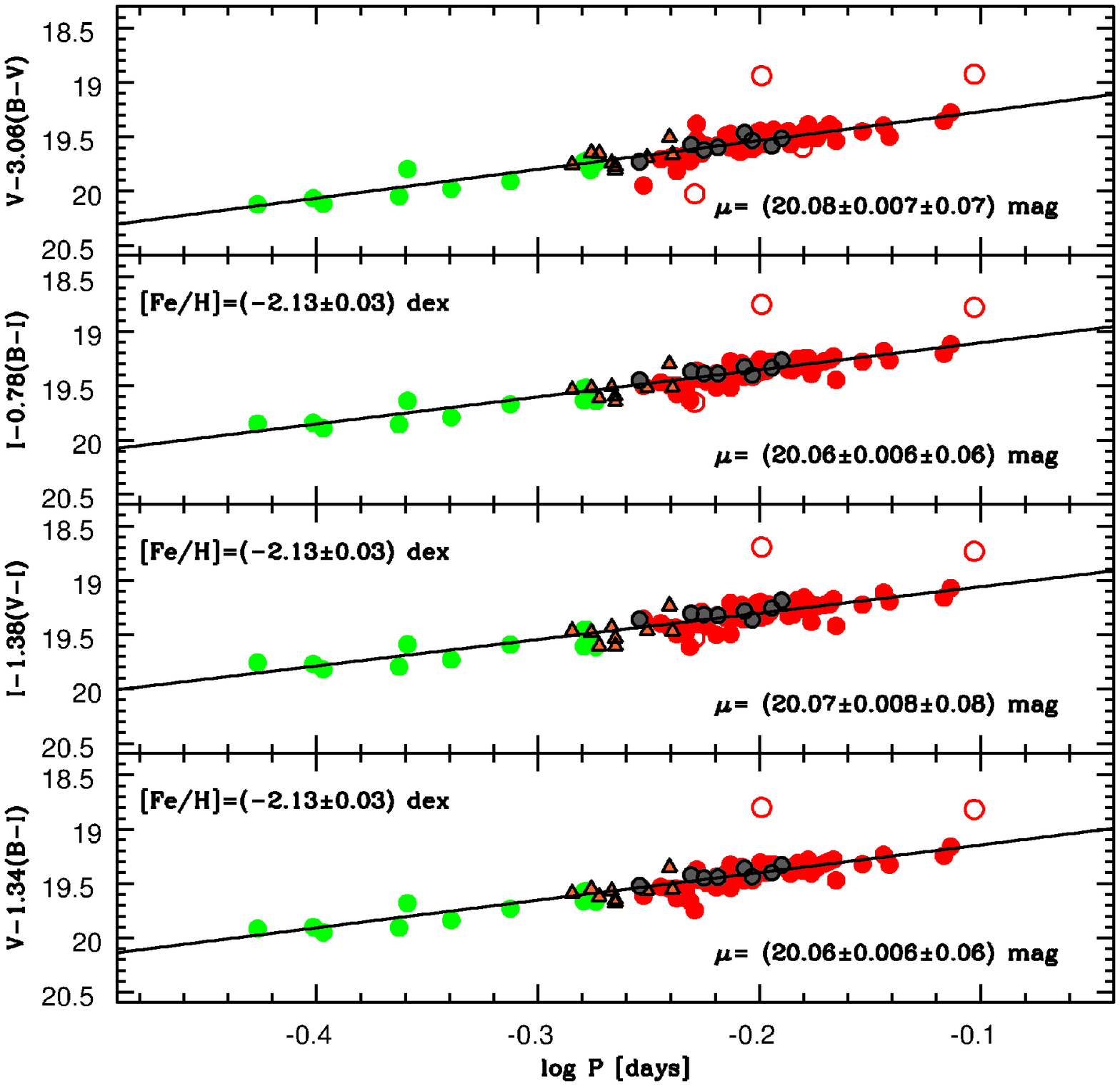} 
\caption{Top -- Predicted PW ({\it BV}) relation (black line). The true distance 
modulus, the standard error of the mean and the standard deviations are labeled. 
Note that this PW relation is independent of the metal content.
Middle -- Same as the top, but for the PW ({\it BI}) relation. The true
distance modulus was estimated using the labeled value of iron abundance.
Bottom -- Same as the middle, but for the PW ({\it VI}) relation.}\label{fig:plw1} 
\end{figure*}

To take advantage of the multiband photometry for Carina RRLs, we
estimated the distance using the PWZ ({\it BI\/}), the PWZ ({\it
  VI\/}) and the triple bands PWZ ({\it BIV\/}) relations (see
Figure~\ref{fig:plw1}). The current pulsation
predictions suggest a mild dependence on the metal content. Indeed, the
coefficients of the metallicity term are $0.106$ ({\it BI\/}), $0.150$
({\it VI\/}) and $0.075$ ({\it BVI\/}). 
In passing we note that the metallicity dependence of the optical PW
relation for RRL stars displays a different trend when compared with
similar relations for classical Cepheids. For the latter objects
the PW ({\it VI}) relation is almost independent of the metallicity while the
PW ({\it BV}) shows a strong dependence on the iron content. The reader
interested in a detailed discussion of the difference between RRLs and
classical Cepheids is referred to the recent paper by M2015. 

To estimate the distance from the PWZ ({\it BI}), the PWZ ({\it VI}) and
the PWZ ({\it BVI}) we adopted
a mean iron abundance for Carina RRLs of ($-2.13 \pm 0.03\pm0.28$) dex~\citep{fabrizio15}. Using this value
we found the following true distance moduli:  $\mu=(20.086\pm 0.006 \pm 0.06)$ mag 
({\it BI}), $\mu=(20.07 \pm 0.008 \pm 0.08)$ mag
({\it VI}) and $\mu=(20.06 \pm 0.006 \pm 0.06)$ mag ({\it VBI}). Note
that the uncertainties in the distance modulus account for the
photometric errors, the standard deviation of the theoretical PWZ relations
and the intrinsic spread in iron abundance. Once again the above
distance determinations agree quite well with the empirical distance,
with the distance based on the PW ({\it BV}) and with distances available in
the literature.

\section{SX Phoenicis} 

More that 30 years ago~\citet{niss81} identified the first variable Blue Straggler (BS)  
in the Galactic globular $\omega$ Cen. The oscillating BSs have been the cross-road 
of several theoretical~\citep{gilliland98,sant01,fiorentino15b} and
observational~\citep{kaluzny03,olech2005,fiorentino14} 
investigations. However, no general consensus has been reached yet on 
their evolutionary and pulsation properties. The names suggested in the literature 
range from SX Phoenicis~\citep{nem90} to Dwarf
Cepheids~\citep{mateo98,vivas13} to ultra-short-period Cepheids~\citep{eggen79} to AI Velorum 
stars~\citep{bessel69} to high-amplitude $\delta$ Scuti stars~\citep{mcnam95,hog97}.

The most relevant point concerning the classification is that SX Phe are considered 
the metal-poor extension of the classical $\delta$ Scuti stars. Ironically, the 
prototype, SX Phe itself, is a metal-intermediate ([Fe/H]=-1.3 dex;
Hog \& Petersen, 1997).  
The circumstantial empirical evidence concerning the pulsation properties are: 
they oscillate both in radial and non radial modes and a significant fraction are 
mixed-mode variables~\citep{gilliland98}; their amplitude ranges from a few 
hundredths to a few tenths of magnitude; they do not show a clear separation in 
the Bailey diagram (luminosity amplitude versus pulsation period), this means 
that the mode identification based on their pulsation properties is not trivial; 
they obey to a period-luminosity~\citep{mcnam00} and to a 
period-luminosity-color-metallicity~\citep{petersen99, fiorentino13}.  

The region of the color-magnitude diagram in which these objects are located 
is strongly affected by degeneracy. They are at the transition between 
low- and intermediate-mass stars during either central hydrogen burning or 
thick shell hydrogen burning. However, the same region is also crossed by 
stellar structures approaching the main sequence \citep{marconi2000} and 
by stellar structures that experienced either a collisional merging or a 
binary merging, i.e. the so-called BSs \citep{dalessandro2013}.

The above evidence indicates that the evolutionary channel producing field 
$\delta$ Scuti and SX Phe can hardly be constrained by their position in the 
Color-Magnitude-Diagram or the Bailey diagram. The empirical scenario becomes 
easier for variables hosted either in open or in globular clusters, since they 
are typically brighter and bluer than MS turn-off stars. This further supporting 
their "peculiar" evolutionary origin.     

In this context, dwarf galaxies play a crucial role. The stellar content of 
dwarf spheroidal galaxies that experienced only a single star formation
event~\citep[Cetus][]{monelli12} and~\citep[Tucana][]{monelli10b}
appears similar to globular clusters. These stellar systems are dominated 
by old (t$>$10 Gyr) stellar populations therefore the color of their main 
sequence turn off stars (MSTO) are systematically redder (cooler) than the 
red edge of the instability strip. This means that the Cepheid instability 
strip in these stellar systems can only be crossed by BSs, since these 
objects are hotter and brighter than canonical MSTO stars. 
However, dwarf galaxies 
that underwent multiple star formation episodes and in particular a well defined 
star formation episode 6-9 Gyrs ago is going to have both "canonical" and "peculiar" 
objects crossing the Cepheid instability strip. Carina dwarf spheroidal belongs to 
the latter group. 

The above scenario has been soundly supported by the recent detailed photometric 
investigation by VM13. They identified more than 340 new 
SX Phe in Carina with period ranging from 0.03860 to 0.18058 days and luminosity amplitudes 
ranging from 0.22 to 1.10 mag in V-band. In this context it is worth mentioning that their 
magnitude (21.89 $\le$ V $\le$ 23.55) and color distribution 
(0.05 $\le$ B-V $\le$ 0.54)  indicate that only a minor fraction belongs 
to the so-called Blue Plume (22.0 $\le$ V $\le$ 23.2; 
0.02 $\le$ B-V $\le$ 0.22), i.e. to the objects for which it is not clear 
whether they are truly young (t $<$ 1 Gyr) or BS of the old populations 
\citep{okamoto2008,monelli12b}. The bulk seems to belong to 
canonical main sequence intermediate-age stars. This evidence opens a new path 
in the investigation of these interesting objects, since we are dealing with 
objects that are the aftermath of the different channels located at the same 
distance and covering a narrow range in metal abundances~\citep{fabrizio15}.        

\begin{figure*}
\includegraphics[scale=0.85]{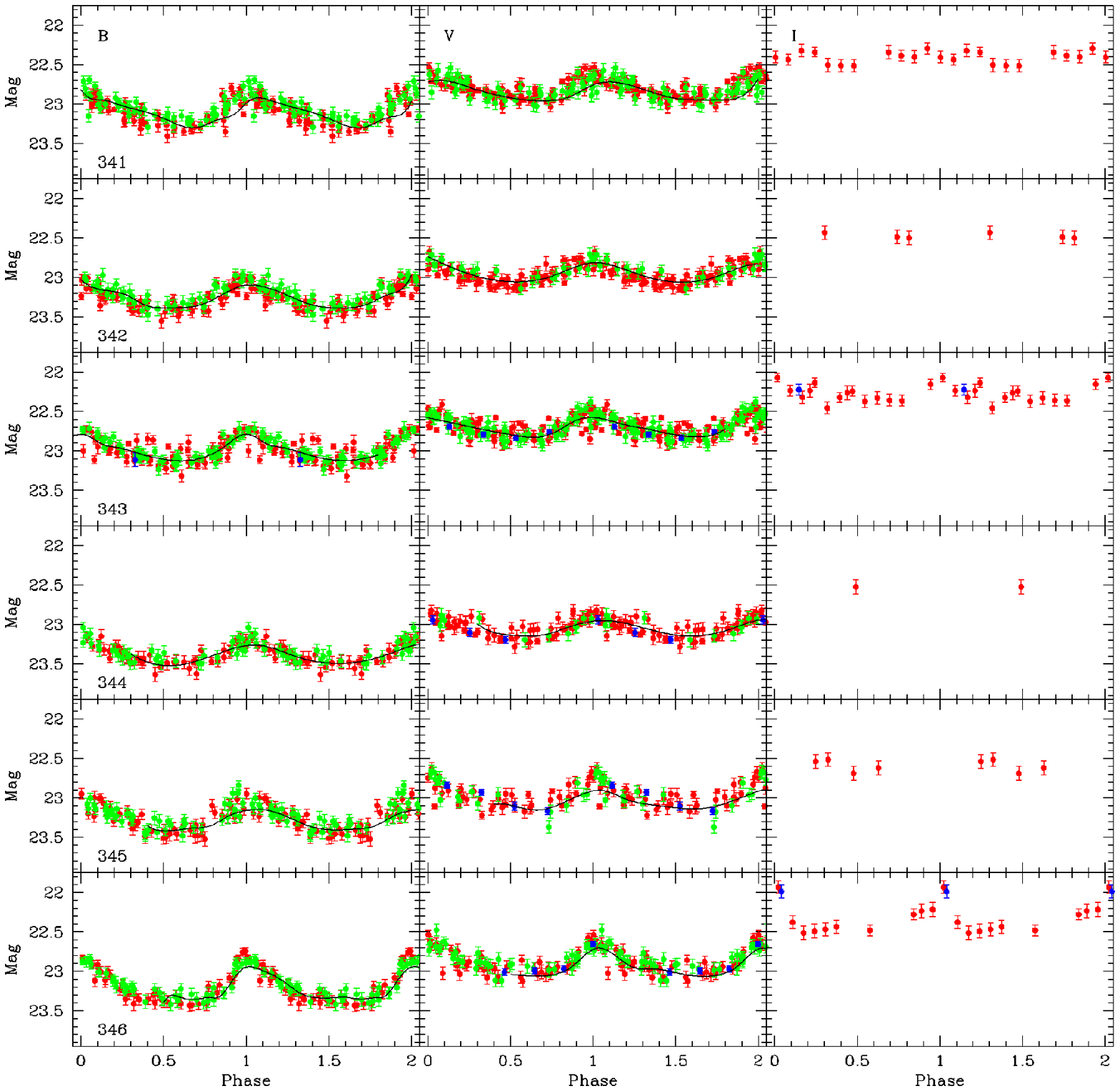} 
\caption{An example of {\it BVI} light curves for selected Carina SX Phe
  variables. Colors are the same of Figure~\ref{fig:lc}. The {\it BVI}  light curves 
for the the entire sample of SX Phe variables are given in a forthcoming paper.}\label{fig_sx_cdl} 
\end{figure*}

Although, the investigation by VM13 is a substantial step forward in the 
identification of these objects we decided to further investigate the 
possible occurrence of SX Phe stars. The working hypothesis was mainly 
supported by the slope of the Cepheid instability strip suggesting that the 
actual sample of SX Phe might be even larger. The similarity of the 
Cepheid instability strip with the location of SX Phe, $\delta$ Scuti,  
and RR Lyrae stars is supported by detailed investigations concerning 
their pulsation properties~\citep{mcnamara11,mcnamara14}.
Thanks to the 
photometric precision and accuracy of our multiband photometric catalog we identified 
101 new SX Phe. Some example of light curves are plotted in Figure~\ref{fig_sx_cdl}, while their
positions, epochs, periods and
mean magnitudes are listed in Table~\ref{tab_sx_new}. The mean magnitudes were estimated as 
intensity means using an analytical fit of the light-curves. Moreover, we confirm the variability 
for 324 out of 
the 340 known SX Phe in Carina. For the others 16, we have
insufficient data  for DC-1, DC-2, DC-3, DC-4, DC-5, DC-6, DC-284 and
DC-339, we do not confirm the variability for DC-75, DC-158, DC-264 and
DC-295, we do not find DC-180 and DC-340 and we consider DC-1, DC-111
and DC-144 the same variable. The pulsation properties for these
objects are given in Table~\ref{tab_sx_old}. 

The comparison between predicted and observed evolutionary and
pulsation properties of Carina SX Phe will be discussed in a forthcoming paper.

%_______________________________________________________________________________
\section{Conclusions and future perspectives}
We have discussed new and accurate multi-band optical ({\it UBVI\/}) photometry of helium burning 
variables in the Carina dwarf spheroidal galaxy. The current photometry covers a time interval 
of more than 20 years. This means that we can provide robust identification of
regular variables with periods close to half and one day, as well as variables 
showing modulations in the pulsation period and/or in in amplitude
(mixed mode pulsators, Blazhko).  

We ended up with a sample of 92 RRLs, among them twelve are first overtones (RR$_c$), 
63 are fundamental (RR$_{ab}$) pulsators, nine mixed-mode variables (RR$_d$) and eight candidate Blazhko variables (Bl).   
Six out of the 92 RRL variables are new identifications. Moreover, we identified 
one new double mode pulsator. 
We also identified 20 ACs, among them one is new identification. 
Together with all these variables we found two new LPVs and seven EB candidates.

For the entire sample of variables we provide accurate pulsation properties 
(periods, luminosity amplitudes) plus accurate estimates of the mean {\it BVI\/} 
magnitudes. 
The mean $BV$ magnitudes are based on a spline 
fit, while the mean $I$-band mean magnitude is based on a template fit. For the RR$_d$
variables we have been able to estimate the two oscillating frequencies and
also the luminosity amplitude of both the FO and the F component. The 
current data do not allow us to constrain the secondary oscillation of the 
candidate Blazhko RRLs. According to extensive photometric surveys of field 
RRLs, they are a minor fraction (8\%) of  Blazhko RRLs~\citep{soszynski11}.  
  
Although the pulsation properties of ACs in Carina are very 
accurate we are still facing the problem of mode identification. 
It seems that optical bands do not help us in settling this
longstanding problem.  

The analysis of the Bailey diagram confirms that Carina is an Oosterhoff 
intermediate system and shows that the luminosity amplitudes of the FO component 
in RR$_{d}$ variables are located along the long--period tail of "regular" RR$_{c}$
variables, while the fundamental components are located along the short--period 
tale of "regular" RR$_{ab}$ variables. To our knowledge this is the first time that 
we can properly locate the two components of RR$_{d}$ variables.

The comparison between theory and observations in the Petersen
diagram for RR$_{d}$ variables indicates that a 
steady increase in the metal content causes a steady decrease in 
the period ratio and in the fundamental period. This evidence is 
supported not only by RR$_{d}$ variables in Galactic globulars, but 
also by RR$_{d}$ variables in the Galactic halo and bulge.   
Moreover, the same diagram shows that RR$_{d}$ variables in nearby
dwarf spheroidals and dwarf irregulars 
(the Magellanic Clouds) display similar properties and that Carina 
RR$_{d}$ variables are located in a region
in which are found only RR$_{d}$ variables hosted in metal-poor 
globular clusters (M15, M68), in the Halo, or in metal-poor dwarf
spheroidal galaxies (e.g. Draco).

The new accurate and precise mean magnitudes allowed us to provide new 
independent estimates of Carina's true distance modulus. We investigated four
different reddening-free Period-Wesenheit relations 
({\it BV\/}, {\it BI\/}, {\it VI\/}, {\it BVI\/}). 
We found that the PW ({\it BV\/}) is independent of the metal content.    
This finding soundly supports recent pulsation predictions based on nonlinear, 
convective, hydrodynamical models of RRL stars (M2015). 

We took advantage of the trigonometric parallaxes for five field RRLs provided 
by~\citet{benedict11} to give a new independent estimate of Carina's
true distance modulus using the observed slope of the PW ({\it BV\/}) relation. We 
found $\mu$=20.02$\pm$0.02 (standard error of the mean) $\pm$0.05 (standard 
deviation) mag. The distance was evaluated using the entire sample of 
variables. In particular, the RR$_c$ variables were fundamentalized. 
The above estimate agrees, within the errors, with Carina distances  
available in the literature that are based on solid standard candles
(see Table~\ref{tab_car_distance}).   

To take advantage of the new predicted optical and NIR PW relations 
provided by M2015 we also estimated the Carina distances 
using the zero-point and the slope of the PWZ ({\it BV\/}, {\it BI\/},
{\it VI\/}, {\it BVI\/})  relations.  
We found true distance moduli of 
$\mu$=(20.08$\pm$0.007$\pm$0.07),  $\mu$=(20.06$\pm$0.006$\pm$0.06),
$\mu$=(20.07$\pm$0.008$\pm$0.08) mag and
$\mu$=(20.06$\pm$0.006$\pm$0.06) mag. 
Note that the distances based on both 
predicted and empirical PW ({\it BV\/}) relations are independent of the metal 
content. The distances based on the  PWZ ({\it BI\/}, {\it VI\/}, {\it BVI\/}) relations have been 
estimated by assuming a mean iron abundance for Carina RRLs of 
[Fe/H]=(--2.13$\pm$0.03$\pm$0.28) dex.  All the above distances are independent of 
uncertainties in the reddening. However, they rely on the assumption that 
the reddening law adopted to estimate the color coefficients of the PW relations,     
is appropriate. The true distances based on empirical and predicted PW relations agree 
quite well one with each other and with similar distances available in
the literature.  

There is evidence that distances based on the theoretical calibrations
are $\sim$0.05-0.1 mag larger the distance based on empirical
calibrations. The evidence applies not only to the PW relation that is
independent of metallicity, but also to the PW relation based on triple bands
indicates that the difference is mainly caused by
a difference in the zero-point. The comparison between distances based
on empirical and on predicted PW relations for stellar systems with precise
distances are required to constrain possible systematics either in the current 
trigonometric parallaxes or in the predicted luminosities at fixed mass.

The above findings further support the mature phase that RRLs are approaching 
as distance indicators. This circumstantial evidence is going to be reinforced 
using either NIR and/or MIR PW relations~\citep{madore13, braga15}.   
The key advantage of the above findings is that RRL PW relations display either 
a mild or a vanishing metallicity dependence. This means the unique opportunity 
to estimate individual distances for a significant fraction of Halo 
and Bulge RRLs that have been recently discovered by long--term photometric 
surveys (OGLE IV,~\citet{soszynski14}; Catalina,~\citet{drake09}; LINEAR,~\citet{palaversa13}; 
LONEOS,~\citet{miceli08}; ASAS,~\citet{pojmanski02};
QUEST,~\citet{vivas04}; Gaia,~\citet{eyer12}). 
This appears as a crucial stepping stone before RRLs can be used as solid 
primary distance indicators in the local Universe.

The recent findings concerning the difference in iron and in Mg distribution 
between the old and the intermediate-age Carina stellar populations opened the 
path for a more detailed pulsation analysis of Carina SX Phe stars. In a recent 
investigation~\citet{fiorentino15b} found that the pulsation properties of 
SX Phe can provide solid constraints on their pulsation masses. This approach 
has already been applied to canonical Galactic globulars
(NGC~6541,~\citet{fiorentino14}) and to Omega Centauri, i.e. in stellar systems 
mainly hosting old stellar populations. The sizable sample of Carina SX Phe
stars is a very promising laboratory to constrain the mass distribution of 
the different stellar populations. Moreover they are located at the same 
distance, and therefore they are an unique opportunity to constrain whether the 
aftermaths of single and binary star evolution display different mass 
luminosity relations. This means the prodrome for their use as solid 
distance indicators.

\acknowledgments   
It is a real pleasure to thank the referee, Mario Mateo, for the
positive opinion concerning the content of the paper and for several
suggestions and insights that improved the readability and the cut of the paper.
This work was partially supported by PRIN--INAF 2011 ``Tracing the formation
and evolution of the Galactic halo with VST'' (PI: M. Marconi) and by
PRIN--MIUR (2010LY5N2T) ``Chemical and dynamical evolution of the Milky Way
and Local Group galaxies'' (PI: F. Matteucci). M. Monelli was supported 
by the Education and Science Ministry of Spain (grants AYA2010-16717).
M. Fabrizio acknowledges financial support from the PO FSE Abruzzo
2007-2013 through the grant “Spectro-photometric characterization of
stellar populations in Local Group dwarf galaxies,”
prot.89/2014/OACTe/D (PI: S. Cassisi). G. Fiorentino was supported by
the Futuro in Ricerca 2013 (RBFR13J716)

%%%%%%%%%%%%%%%%%%%%%%%%%%%%%%%%%%%%%%%%%%%%%%%%%%%%%%%%%%%%%%%%%%%%%d
%			Table 
%%%%%%%%%%%%%%%%%%%%%%%%%%%%%%%%%%%%%%%%%%%%%%%%%%%%%%%%%%%%%%%%%%%%%d

\clearpage
\begin{table}
\begin{center}
\caption{Log of observations.\label{tab_obs}}
\begin{tabular}{lllllcccccc}
\tableline
\tableline
& Run ID & Dates & Telescope &  Camera & $U$ & $B$ & $V$ & $R$ & $I$ & Multiplex \\
\tableline
1 & susi9810:     &  1998 10 26-29          & ESO NTT 3.6m    & SUSI  & -- & -- & 2 & -- & 3 &     $\times 2$          \\
  2 & wfi38:     &  1999 10 30-11 14             & MPI/ESO 2.2m & WFI      & -- & 3 &  -- & 2 & -- &  $\times 8$   \\
  3 & susi0410: &  2004 10 05-09            & ESO NTT 3.6m & SUSI    & -- & -- &  
 -- & 12 &  -- &   $\times 2$            \\
  4 & B05jan12:    &  2005 01 12          & CTIO 4.0m & Mosaic2      & 12 & -- & -- & -- & 5 &  $\times 8$   \\
  5 & B05jan17:   &  2005 01 18 & CTIO 4.0m    & Mosaic2  & -- & 2 & 6 & -- & 21 &  $\times 8$   \\
  6 & B0512:    &   2005 12 28  & CTIO 4.0m & Mosaic2      & 14 & -- & 2 & -- &  1 &  $\times 8$    \\
  7 & B0712:   &  2007 12 14-19          & CTIO 4.0m & Mosaic2      & -- & 71 & 88 & -- & -- &  $\times 8$    \\
  8 & B1002:    &  2010 01 14-16            & CTIO 4m & Mosaic2      & -- & -- & 19 & -- & -- &  $\times 8$    \\
  9 & wfi41:    &  2012 02 22-29             & MPI/ESO 2.2m & WFI      & 6 &  3 &  -- & -- &  3 &  $\times 8$    \\
\tableline
\end{tabular}       
\end{center}
\end{table}

\clearpage
\LongTables
\begin{deluxetable*}{lllclllc}
\tabletypesize{\scriptsize}
\tablecolumns{16}
\tablewidth{500pt}
% \tabletypesize{\footnotesize}
% \tablecolumns{10}
\tablecaption{Catalog of Cartina variable stars. \textit{Col.} $1$: identification star;
\textit{Col.} $2$: right ascension, ($\alpha$); \textit{Col.} $3$: declination, ($\delta$); \textit{Col.} $4$: type as in S86;
\textit{Col.} $5$: type and identification  as in D03; \textit{Col.} $6$: type and identification as in VM13; \textit{Col.} $7$: identification and type as in Paper VI;\textit{Col.} $8$: type as in this study.\label{tab_coordinates}}
\tablehead{\colhead{ID}  & \colhead{$\alpha$ (J$2000.0$)} & \colhead{$\delta$ (J$2000.0$)} & \colhead{S$98$}  &\colhead{MDO$03$}  & \colhead{VM$13$} &\colhead{Paper I} &\colhead{Our} \\ 
\nodata 	&	(hh mm ss)         & (dd mm ss) &\nodata	& \nodata&\nodata& \nodata&\nodata}
 \startdata
V3     &      06 43 31.71  &  --50 51 05.3 &      ab    &  \dots& \ldots &  V215 (ab) & ab \\
V4     &      06 43 28.70  &  --50 55 46.9 &      c      &  V177 (c)& \ldots& \ldots&c     \\
V7     &      06 43 16.26  &  --51 04 17.2 &      ab    &  V7 (ab)  &   \ldots&V7 (ab) &ab \\          
V10   &      06 43 10.66  &  --50 43 47.8 &      ab    &  V10 (ab) & \ldots &V10 (ab) & ab\\	     
V11   &      06 43 08.20  &  --50 50 47.9 &      c      &  V11 (d)  & \ldots& V11 (d)   & d   \\        
V14   &      06 43 01.93  &  --51 01 26.3 &      ab    &   V14 (AC) &  \ldots& V14 (AC) & AC\\    
V22   &      06 42 46.32  &  --50 57 36.9 &  	 ab     &   V22 (c)   & \ldots & V22 (ab) & ab\\     
V24   &      06 42 34.00  &  --50 44 24.9 &      ab    &   V24 (ab)  &\ldots&V24 (ab)   & ab \\
\enddata
\tablenotetext{FV}{Field variable.}
\tablenotetext{RGB}{Variable located along the RGB.}
\tablenotetext{SV}{Suspected variable.}
 \end{deluxetable*}

\clearpage
\LongTables
\begin{deluxetable*}{llllll}
\tablewidth{0pt}
\tablecaption{Catalog of candidate variable stars that according to the current investigation are not variable. \textit{Col.} $1$: identification star;
 \textit{Col.} $2$: right ascension, ($\alpha$); \textit{Col.} $3$: declination, ($\delta$); \textit{Col.} $4$: type as in S86;
 \textit{Col.} $5$: type as in D03;  \textit{Col.} $6$: type as in this study.\label{tab_nvcoordinates}}
\tablehead{\colhead{ID}  & \colhead{$\alpha$ (J$2000.0$)} & \colhead{$\delta$ (J$2000.0$)} & \colhead{S$86$}  &\colhead{MDO$03$}  & \colhead{This Work} \\ 
\nodata	&	(hh mm ss)         & (dd mm ss) &\nodata	& \nodata& \nodata}
\startdata
V1   &      06 42 21.90  & --50 51 35.0  & ab  &\ldots&NV\\
V2   &      06 42 19.60  & --50 45 12.0  & SV  &\ldots&NV\\
V5   &	06 43 19.04  & --50 53 29.3  & SV  &         NV	  	&	NV	\\  	
V6   &      06 43 18.01  &  --50 46 43.7 &  SV  &        NV		&	NV	        \\
V8   &      06 42 56.44  &  --50 41 45.6 &$\delta$Scu&         NV		&	NV	\\       
V9   &      06 43 16.00  &  --51 02 43.6 &     WUma&    NV		&	NV$^{b}$ \\ 
V12 &      06 43 06.20  &  --50 59 04.6 &     $\delta$Scu&    NV		&	NV$^{c}$\\	 
V13 &      06 43 05.02  &  --50 49 33.0 & SV&        NV		&	NV$^{c}$   \\
\enddata
 \scriptsize
\tablenotetext{NV}{No variability detected.}
\tablenotetext{SV}{Suspected variable.}
\tablenotetext{a}{Possible field galaxy.}
\tablenotetext{b}{Possible blend.}
\tablenotetext{c}{Located close to a bright or faint star.}
\tablenotetext{d}{Saturated star.}
\tablenotetext{e}{Faint star.}
\end{deluxetable*}

\clearpage

\begin{table}
\begin{center}
\caption{Photometry of the Carina variable stars. \label{tab_phot}}
\begin{tabular}{llllllllllll}
\tableline
\tableline
% \tabletypesize{\footnotesize}
HJD&U & $\sigma_{U}$ & HJD&  B & $\sigma_{B}$ &HJD& V & $\sigma_{V}$&HJD& I & $\sigma_{I}$\\
\tableline
V7                    &             &           & & &     &                       &             &              &                     &             &           \\   
\tableline
53351.7001&	21.50&	0.03&    51552.5736&	21.33&	0.02&	  51552.5669&	20.90&	0.02&	  53351.7698&	20.38&	0.02\\	
53351.7129&	21.64&	0.03&    51552.5854&	21.40&	0.02&	  51552.5786&	20.90&	0.01&	  53351.7745&	20.40&	0.02	\\
53351.7256&	21.59&	0.03&    51552.5971&	21.43&	0.02&	  51552.5904&	20.94&	0.02&	  53351.7791&	20.38&	0.02	\\
53351.7387&	21.59&	0.04&    51552.6089&	21.43&	0.02&	  51552.6021&	20.92&	0.01&	  53351.7838&	20.37&	0.02	\\
53351.7515&  21.62&	0.04&    51552.6209&	21.47&	0.01&	  51552.6141&	21.03&	0.01&	  53351.7885&	20.43&	0.02	\\
53351.7642&  21.68&	0.03&    51552.6327&	21.49&	0.02&	  51552.6259&	20.99&	0.02&	  53351.7931&	20.35&	0.02	\\
53359.7729&  21.39&	0.03&    51552.6445&	21.53&	0.02&	  51552.6378&	21.04&	0.01&	  53351.7978&	20.43&	0.02	\\
53359.7858&  21.14&	0.02&    51552.6562&	21.53&	0.02&	  51552.6495&	21.04&    0.02&	  53359.6762&	20.39&	0.01	\\
53359.7992&  20.75&	0.02&    51552.6681&	21.56&	0.02&	  51552.6613&	21.08&	0.01&	  53359.6808&	20.39&	0.02	\\
53359.8121&  20.45&	0.02&    51552.6798&	21.54&	0.02&	  51552.6731&	21.04&	0.02&	  53359.6855&	20.34&	0.02 \\
                   &           &            &    51552.6916&	21.58&	0.02&        51552.6849&	        21.10&	0.02&         53359.6901&	20.36&	0.01 \\
\tableline
\end{tabular}       
\end{center}
\end{table}

\clearpage
\LongTables
\begin{deluxetable*}{l c l l  c c c c c c c c}
\tabletypesize{\scriptsize}
\tablecolumns{16}
\tablewidth{500pt}
\tablecaption{Pulsation Properties of Carina variable stars.\label{tab_prop}}
\tablehead{
\colhead{ID} & \colhead{Type} &\colhead{Epoch} &\colhead{Period (d)} & \colhead{$\langle U \rangle^{b}$} & \colhead{$\langle B \rangle^{b}$} & \colhead{$\langle V \rangle^{b}$} &\colhead{$\langle I \rangle^{b}$} & \colhead{A($U$)} & \colhead{A($B$)} & \colhead{A($V$)}& \colhead{A($I$)}}
\startdata
3 & 	ab  &  	   52351.2771  &    0.5786581  & 21.30$^{c}$  & 	  21.127&          20.771&    20.210& \ldots& 1.256& 1.012&  0.637   \\ 
4 &   c   &            51548.6519  &    0.3973509  & 21.243 &    20.995&          20.685&    20.234& 0.513& 0.663& 0.565&  0.356   \\ 	  
7    & 	ab  &  	51548.7025  &    0.6033121  & 21.43$^{c}$  &  21.104&  20.729&  20.212&  \ldots&  1.278&  1.087&  0.685\\ 
10  &	ab  &  	51548.7378  &    0.5845143  & 21.158         &  21.070&  20.724&  20.203&   0.712&  1.367&  1.095&  0.690\\ 	  	
14  &	AC &        51548.9240  &    0.4766975  & 20.510         &  20.265&  19.994&  19.507&   0.838&  0.886&  0.745&  0.469\\ 
17  &       EB  &        51548.5693  &    0.3933393  & 22.03$^{c}$  &  20.968&  19.606&  17.705&   \ldots&  0.092&  0.086&  0.059\\ 
22  &	ab  &  	51549.0012  &    0.6380630  & 21.209         &  21.155&  20.746&  20.133&   0.757&  0.709&   0.518& 0.326\\ 	
24  & 	ab  &  	51548.5844  &    0.6181987  & 21.48$^{c}$  &  21.088&   20.730&    20.126& \ldots& 0.909& 0.723&	0.456	 \\
\enddata
\tablenotetext{a}{Epoch of maximum light.}
\tablenotetext{b}{Intensity--averaged magnitudes.}
\tablenotetext{c}{Median of individual magnitude measurements.}
\tablenotetext{d}{Not enough data to fit a light curve, uncertain parameters.}
\end{deluxetable*}

\clearpage

\begin{turnpage}
\begin{deluxetable}{llllccccccccc}
\tablewidth{500pt}
\tabletypesize{\scriptsize}
\tablecaption{Pulsation properties of Carina RR$_{d}$ variable stars. \label{tab_prop_rrd}}
\tablehead{ID & Epoch &P$_{0}$ & P$_{1}$/P$_{0}$ & $\langle B \rangle$ & $\langle V \rangle$ & $\langle I \rangle$ & A($B_0$) & A($B_1$) & A($V_0$)& A($V_1$) & A ($I_0$)& A($I_1$) \\
   ~~~~& ~~~~&  (days) & (days)& (mag)	&(mag)& (mag) & (mag)	&(mag)& (mag) & (mag)	&(mag)& (mag)}
\startdata
V11     &51548.9191 &  0.543689  &0.745854	& 21.080&  20.766 &  20.247&0.548&0.341&0.451&0.339&0.28&0.21\\
V26     &51548.9291 &  0.562177  &0.745703	& 20.990&  20.670 &  20.162&0.604&0.488&0.610&0.395 &0.39&0.25\\
V74     &51548.7140 &  0.533702  &0.747205       & 21.059&  20.712 &  20.246&0.607&0.222&0.567& 0.263&0.36&0.17\\
V89     &51548.6377 &  0.519403  &0.746522       &21.102 & 20.770&   20.222&0.408&0.384&0.524&0.278&0.33&0.17\\
V192   &51549.0958 &  0.541694  &0.748661      &21.030&  20.720&    20.252&0.753&0.525&0.600&0.345&0.38&0.22\\ 
V198   &51548.6238 &  0.530551   &0.745534   & 21.058 & 20.710&    20.192&0.664&0.323&0.506&0.334&0.32&0.21\\
V207   &51549.1028 &  0.541156 &0.746387	 & 21.122 &20.782 &   20.215&0.621&0.548&0.637&0.349&0.40&0.22\\
V210   &52351.5738 & 0.57324    &0.75286  &21.010&    20.639&    20.048&0.474&0.275&0.418&0.218&0.26&0.14\\
V225   &52351.8655 & 0.57688    &0.74610  &20.960&    20.641&    20.147&0.776&0.363&0.575&0.324&0.36&0.20\\ 
\enddata
\end{deluxetable}
\end{turnpage}

\clearpage

\begin{deluxetable}{lcccccc}
\tablewidth{500pt}
\tabletypesize{\scriptsize}
\tablecaption{Pulsation properties for the nine extra--tidal variables (six RRLs, four ACs)
identified by VM13 \label{tab_prop_vivas}}
\tablehead{ID &Type & P & (B) & (V) & A($B$) & A($V$)\\
   ~~~~& ~~~~&  (days) & (mag)	& (mag) & (mag)& (mag)}
\startdata
RRL--1    &ab & 0.629&20.94&20.55&0.55&0.45\\
RRL--2     &ab&0.523&21.04&20.60&0.83&0.65\\
RRL--3     &ab&0.544&21.06&20.62&0.56&0.46\\
RRL--4     &c& 0.204&21.32&20.85&0.17&0.24\\
RRL--5     &c&0.107&21.27&20.82&0.14&0.27\\
RRL--38   &c&0.189&21.18&20.77&0.41&0.31\\
AC--1 &\ldots&0.186&20.44& 19.95& 0.21 & 0.20\\
AC--9&\ldots& 0.476&19.21& 19.13& 0.36 & 0.55\\
AC--10&\ldots& 0.163&19.71& 19.35& 0.18 & 0.32\\
\enddata
\end{deluxetable}

\clearpage

\begin{deluxetable}{lclll}
\tablewidth{0pt}
\tabletypesize{\scriptsize}
\tablecaption{Number of RR$_{d}$ hosted in the stellar systems plotted in Figure~\ref{petersen2} 
and their references. Columns 4 and 5 give the iron abundances of the hosting stellar system and 
their references.\label{tab_rrd_other}}
\tablehead{ System& N & Reference & $[Fe/H]$ & Reference}
\startdata
NGC 2419                    & 1   & \cite{clement93}     &  --2.20$\pm$ 0.09&\cite{carretta09} \\
NGC 6426                    & 1   & \cite{clement93}     &  --2.33$\pm$0.15&\cite{hatzidimitriou99}\\
LMC                         & 985 & \cite{soszynski09}   &  --0.33$\pm$0.13&\cite{romaniello08}$^{a}$\\
SMC                         & 257 & \cite{soszynski10}   &  --0.75$\pm$0.08&\cite{romaniello08}$^{a}$\\
Bulge                       & 173 & \cite{soszynski14}   &  --1.5/--0.5& \cite{zoccali08}$^{b}$\\
Halo (NSV 09295)            & 1   &  \cite{garcia97}     &  --1.5 &\cite{layden94}\\
Halo (AQ Leo)               & 1   & \cite{clement91}     &  --1.5 &\cite{layden94}\\
Halo (VIII--10, VIII--58)     & 2   & \cite{clement93}     &  --1.5 &\cite{layden94} \\
\enddata         
\tablenotetext{a}{Mean iron abundances and standard deviations of classical Cepheids based on high 
spectral resolution spectra. Metal abundances for 98 LMC RRLs was provided by \citet{gratton04} using 
low--resolution spectra and found $[Fe/H]=--1.48 \pm 0.03 \pm 0.06$. Metal abundances for SMC RRLs 
are not available. The iron abundance of the single SMC GC (NGC~121) is [Fe/H]=--1.19$\pm$0.12 
provided by~\citet{dacosta98}.}
\tablenotetext{b}{Range in iron abundance covered by red giants in the Galactic bulge. 
The iron abundance of Bulge RRLs was provided by~\citet{walker91} [Fe/H]=--1.00$\pm$0.16 
using low--resolution spectra.}
\end{deluxetable}           
\clearpage

\begin{deluxetable}{ccc}
\tablewidth{0pt}
\tabletypesize{\scriptsize}
\tablecaption{Results of the empirical fit of Period--Wesenheit relations $W=\alpha+\beta \log P_F$.\label{table:wesenheit1}}
\tablehead{$\alpha$  & $\beta$ & rms} 
\startdata
\multicolumn{3}{c}{$W(B,V)=M_V-3.06(B-V)$}\\

(18.98$\pm$0.03) &(-2.7$\pm$0.1)& 0.07\\
\\
\multicolumn{3}{c}{$W(B,I)=M_I-0.78(B-I)$}\\

(18.84$\pm$0.03) &(-2.6$\pm$0.1) & 0.06\\
\\

\multicolumn{3}{c}{$W(V,I)=M_I-1.38(V-I)$}\\

(18.80$\pm$0.03)& (-2.5$\pm$0.1) &0.08\\
\\
\multicolumn{3}{c}{$W(V,BI)=M_V-1.34(B-I)$}\\

(18.88$\pm$0.03)& (-2.6$\pm$0.1) &0.06\\
\\
\hline
\enddata
\end{deluxetable}

\clearpage

\begin{deluxetable}{lccccl}
\tablewidth{0pt}
\tabletypesize{\scriptsize}
\tablecaption{Periods, $B$ and $V$ mean magnitudes, true distance
  moduli and metallicity for 
the five field RRLs stars by~\citet{benedict11}.\label{tab_calibrating}}
\tablehead{Name&$\log$ P & $<B>$ & $<V>$ & $\mu$ & [Fe/H]}

\startdata
RR Lyr &--0.24655&8.07$\pm$0.01&7.75$\pm$0.01&7.14$\pm$0.07&--1.41$\pm$0.13\\
SU Dra &--0.18018&10.23$\pm$0.01&9.95$\pm$0.01&9.35$\pm$0.24&--1.80$\pm$0.2\\
UV Oct &--0.26552&9.71$\pm$0.01&9.35$\pm$0.01&8.87$\pm$0.13&--1.47$\pm$0.11\\
XZ Cyg &--0.33107&10.13$\pm$0.01&9.84$\pm$0.01&8.98$\pm$0.22&--1.44$\pm$0.2\\
RZ Cep &--0.38352&9.92$\pm$0.01&9.46$\pm$0.01&8.03$\pm$0.16&--1.77$\pm$0.2\\
\enddata         
\tablenotetext{}{The distance moduli in this column are slightly
  different than those in Table~8 of~\citet{benedict11} due to
  typographical errors in the paper.\\
Iron abundances are on the~\citet{zw84} metallicity scale
  according to~\citet{benedict11},  (see their Table~1).}
\end{deluxetable}

\clearpage

\begin{deluxetable}{clll}
\tablewidth{0pt}
\tabletypesize{\scriptsize}
\tablecaption{Carina true distance moduli based on different diagnostics.\label{tab_car_distance}}
\tablehead{ $\mu$& E(B--V)&Method & Reference}
\startdata
$20.06 \pm 0.12$ & 0.025&PL (DC)       &      \cite{mateo98} \\
$20.10 \pm 0.12$ &0.03& PLC & D03 \\
$20.00 \pm 0.10$ & 0.03&PLA & D03 \\
$20.10 \pm 0.04$ & 0.03&FOBE & D03 \\
$20.11 \pm 0.13$ & 0.06&TRGB & \cite{pietrinsky09} \\
$20.17 \pm 0.10$ &0.063& PL (DC) &  \cite{vivas13} \\
$20.09 \pm 0.07$ &0.03& PLW& Paper VI\\
\enddata         
\end{deluxetable}

\clearpage

\LongTables
\begin{deluxetable*}{l l l l  c c c}
\tabletypesize{\scriptsize}
\tablecolumns{16}
\tablewidth{500pt}
\tablecaption{Pulsation Properties of the new Carina SX Phe.\label{tab_sx_new}}
\tablehead{
\colhead{ID} & \colhead{$\alpha$} &\colhead{$\delta$} &\colhead{Period (d)} & \colhead{$\langle B \rangle^{b}$} & \colhead{$\langle V \rangle^{b}$} &\colhead{$\langle I \rangle^{b}$}}
\startdata
341  &  06 40 14.14 & --51 01 23.4 &    0.05898004 &  23.11 &   22.85 &   22.42    \\
342  &  06 40 15.46 & --51 08 31.2 &    0.06404925 &  23.25 &   22.94 &   22.50    \\
343  &  06 40 17.76 & --50 56 06.0 &    0.06189917 &  22.99 &   22.70 &   22.33    \\
344  &  06 40 23.45 & --51 00 18.5 &    0.05700707 &  23.39 &   23.04 &   22.49    \\
345  &  06 40 23.84 & --50 55 16.0 &    0.0616327   &  23.28 &   23.04 &   22.65    \\
346  &  06 40 26.44 & --50 55 12.2 &    0.06983306 &  23.20 &   22.93 &   22.48    \\
347 &  06 40 34.03 & --50 56 36.9 &    0.06382168 &  23.17 &   22.92 &   22.49    \\
348  &  06 40 38.49 & --50 50 34.2 &    0.05887662 &  23.01 &   22.74&   22.34    \\
\enddata
\end{deluxetable*}
\clearpage
\LongTables
\begin{deluxetable*}{l l l l  c c c c}
\tabletypesize{\scriptsize}
\tablecolumns{16}
\tablewidth{500pt}
\tablecaption{Pulsation Properties of the known Carina SX Phe.\label{tab_sx_old}}
\tablehead{
\colhead{ID} & \colhead{$\alpha$} &\colhead{$\delta$} &\colhead{Period (d)} & \colhead{$\langle B \rangle^{b}$} & \colhead{$\langle V \rangle^{b}$} &\colhead{$\langle I \rangle^{b}$} & M98}
\startdata
7  &      06 39 53.11& --51 04 31.5 &    0.05861265  &    23.43 & 23.13 & 22.12 &\ldots  \\ 
8  &      06 39 56.16& --51 06 43.6 &    0.0840233   &    22.66 & 22.36 & 21.92  &\ldots\\ 
9  &      06 40 00.79& --50 57 58.5 &    0.06283058  &    23.33 & 23.04 & 22.45 &\ldots \\ 
10 &      06 40 03.36&  --51 07 55.6 &    0.06600411  &    23.19 & 22.93 & 22.43&\ldots  \\ 
11 &      06 40 06.08&  --50 55 58.0 &    0.0566776   &    23.32 & 23.04 & 22.58 &\ldots \\ 
12 &      06 40 14.68&  --51 05 05.0 &    0.05116040  &    23.46 & 23.18 &\ldots&\ldots\\   
13 &      06 40 17.93&  --50 56 33.7 &    0.0572522   &    23.27 & 23.01 & 22.59 &\ldots  \\
14 &      06 40 18.59&  --51 03 01.9 &    0.05801669  &    23.33 & 23.06 & 22.67&\ldots  \\
\enddata
\end{deluxetable*}
\clearpage

%%%%%%%%%%%%%%%%%%%%%%%%%%%%%%%%%%%%%%%%%%%%%%%%%%%%%%%%%%%%%%%%%%%%%%%%%%%%%%%
%				Fig. 1
%%%%%%%%%%%%%%%%%%%%%%%%%%%%%%%%%%%%%%%%%%%%%%%%%%%%%%%%%%%%%%%%%%%%%%%%%%%%%%%

%%%%%%%%%%%%%%%%%%%%%%%%%%%%%%%%%%%%%%%%%%%%%%%%%%%%%%%%%%%%%%%%%%%%%%%%%%%%%%%
%				Fig. 1BIS
%%%%%%%%%%%%%%%%%%%%%%%%%%%%%%%%%%%%%%%%%%%%%%%%%%%%%%%%%%%%%%%%%%%%%%%%%%%%%%%

%%%%%%%%%%%%%%%%%%%%%%%%%%%%%%%%%%%%%%%%%%%%%%%%%%%%%%%%%%%%%%%%%%%%%%%%%%%%%%%
%				Fig. 2
%%%%%%%%%%%%%%%%%%%%%%%%%%%%%%%%%%%%%%%%%%%%%%%%%%%%%%%%%%%%%%%%%%%%%%%%%%%%%%%

%%%%%%%%%%%%%%%%%%%%%%%%%%%%%%%%%%%%%%%%%%%%%%%%%%%%%%%%%%%%%%%%%%%%%%%%%%%%%%%
%				Fig. 19
%%%%%%%%%%%%%%%%%%%%%%%%%%%%%%%%%%%%%%%%%%%%%%%%%%%%%%%%%%%%%%%%%%%%%%%%%%%%%%%
\begin{figure*}
\includegraphics[scale=0.8]{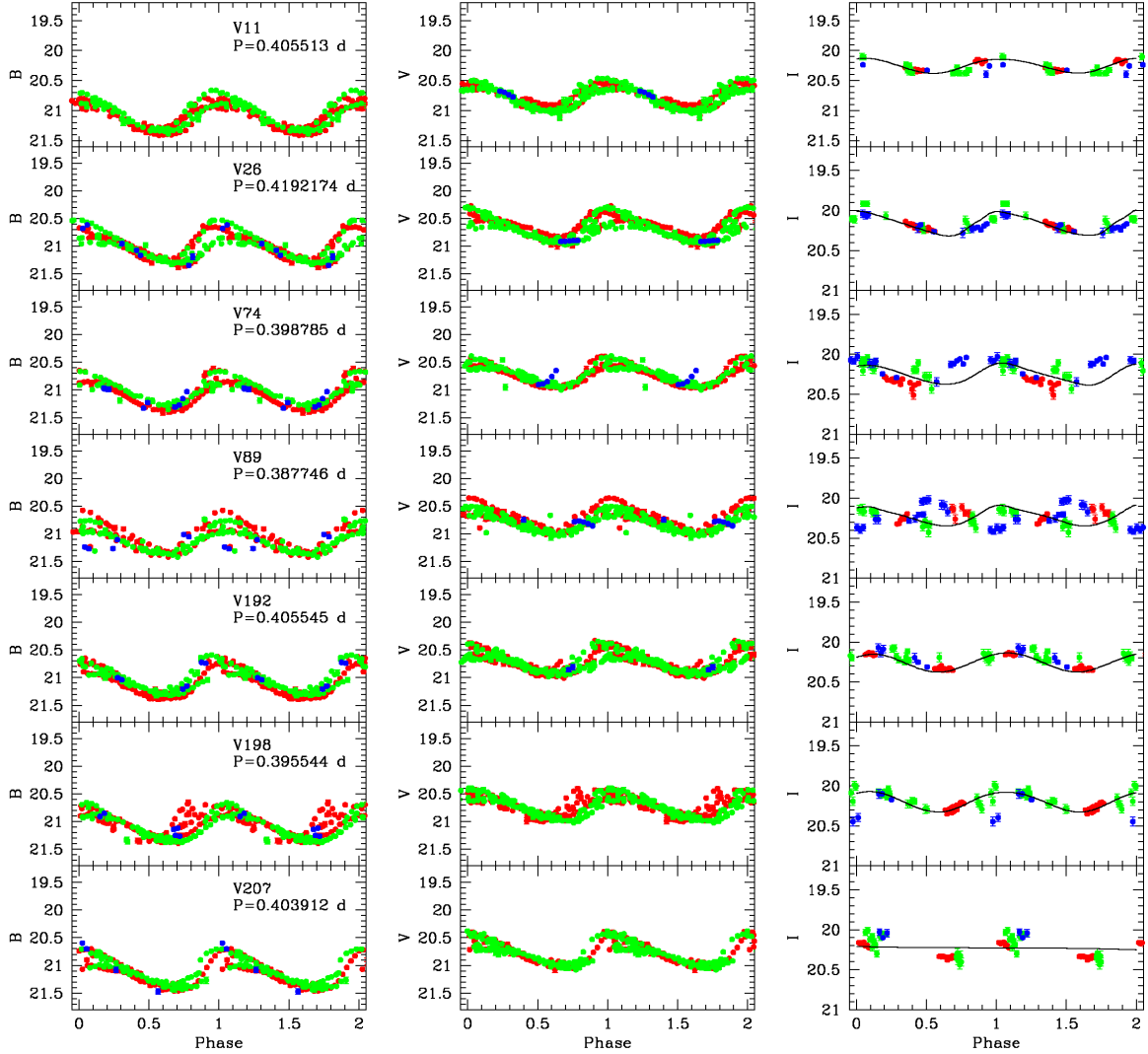} 
\caption{Same as Figure \ref{fig:lc}, but for the BVI light curves of the nine RR$_{d}$ stars in 
our sample.}\label{fig:rrd1}
\end{figure*}

%%%%%%%%%%%%%%%%%%%%%%%%%%%%%%%%%%%%%%%%%%%%%%%%%%%%%%%%%%%%%%%%%%%%%%%%%%%%%%%
%				Fig. 20
%%%%%%%%%%%%%%%%%%%%%%%%%%%%%%%%%%%%%%%%%%%%%%%%%%%%%%%%%%%%%%%%%%%%%%%%%%%%%%%
\begin{figure*}
\includegraphics[scale=0.6]{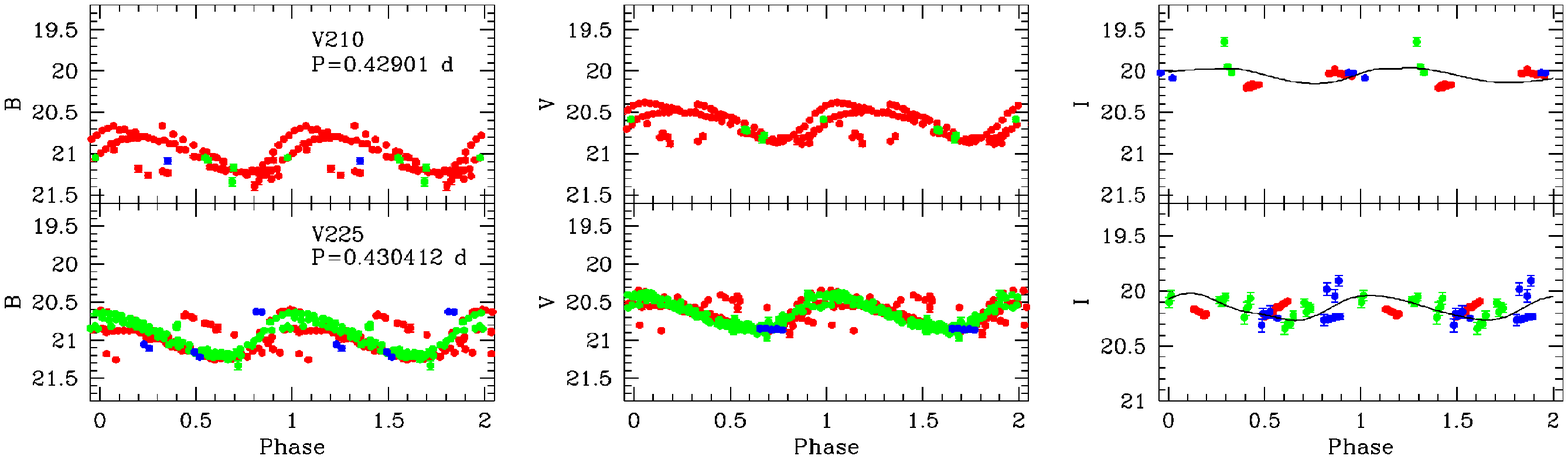} 
\figurenum{\ref{fig:rrd1}}
\caption{(Continued).}
\end{figure*}

%%%%%%%%%%%%%%%%%%%%%%%%%%%%%%%%%%%%%%%%%%%%%%%%%%%%%%%%%%%%%%%%%%%%%%%%%%%%%%%
%				Fig. 3
%%%%%%%%%%%%%%%%%%%%%%%%%%%%%%%%%%%%%%%%%%%%%%%%%%%%%%%%%%%%%%%%%%%%%%%%%%%%%%%

%%%%%%%%%%%%%%%%%%%%%%%%%%%%%%%%%%%%%%%%%%%%%%%%%%%%%%%%%%%%%%%%%%%%%%%%%%%%%%%
%				Fig. 4
%%%%%%%%%%%%%%%%%%%%%%%%%%%%%%%%%%%%%%%%%%%%%%%%%%%%%%%%%%%%%%%%%%%%%%%%%%%%%%%

%%%%%%%%%%%%%%%%%%%%%%%%%%%%%%%%%%%%%%%%%%%%%%%%%%%%%%%%%%%%%%%%%%%%%%%%%%%%%%%
%				Fig. 5
%%%%%%%%%%%%%%%%%%%%%%%%%%%%%%%%%%%%%%%%%%%%%%%%%%%%%%%%%%%%%%%%%%%%%%%%%%%%%%%

%%%%%%%%%%%%%%%%%%%%%%%%%%%%%%%%%%%%%%%%%%%%%%%%%%%%%%%%%%%%%%%%%%%%%%%%%%%%%%%
%				Fig. 6
%%%%%%%%%%%%%%%%%%%%%%%%%%%%%%%%%%%%%%%%%%%%%%%%%%%%%%%%%%%%%%%%%%%%%%%%%%%%%%%

%%%%%%%%%%%%%%%%%%%%%%%%%%%%%%%%%%%%%%%%%%%%%%%%%%%%%%%%%%%%%%%%%%%%%%%%%%%%%%%
%				Fig. 7
%%%%%%%%%%%%%%%%%%%%%%%%%%%%%%%%%%%%%%%%%%%%%%%%%%%%%%%%%%%%%%%%%%%%%%%%%%%%%%%

%%%%%%%%%%%%%%%%%%%%%%%%%%%%%%%%%%%%%%%%%%%%%%%%%%%%%%%%%%%%%%%%%%%%%%%%%%%%%%%
%				Fig. 8
%%%%%%%%%%%%%%%%%%%%%%%%%%%%%%%%%%%%%%%%%%%%%%%%%%%%%%%%%%%%%%%%%%%%%%%%%%%%%%%

%%%%%%%%%%%%%%%%%%%%%%%%%%%%%%%%%%%%%%%%%%%%%%%%%%%%%%%%%%%%%%%%%%%%%%%%%%%%%%%
%				Fig. 9
%%%%%%%%%%%%%%%%%%%%%%%%%%%%%%%%%%%%%%%%%%%%%%%%%%%%%%%%%%%%%%%%%%%%%%%%%%%%%%%

%%%%%%%%%%%%%%%%%%%%%%%%%%%%%%%%%%%%%%%%%%%%%%%%%%%%%%%%%%%%%%%%%%%%%%%%%%%%%%%
%				Fig. 10
%%%%%%%%%%%%%%%%%%%%%%%%%%%%%%%%%%%%%%%%%%%%%%%%%%%%%%%%%%%%%%%%%%%%%%%%%%%%%%%

%%%%%%%%%%%%%%%%%%%%%%%%%%%%%%%%%%%%%%%%%%%%%%%%%%%%%%%%%%%%%%%%%%%%%%%%%%%%%%%
%				Fig. 11
%%%%%%%%%%%%%%%%%%%%%%%%%%%%%%%%%%%%%%%%%%%%%%%%%%%%%%%%%%%%%%%%%%%%%%%%%%%%%%%
\begin{figure*}
\includegraphics[scale=0.85]{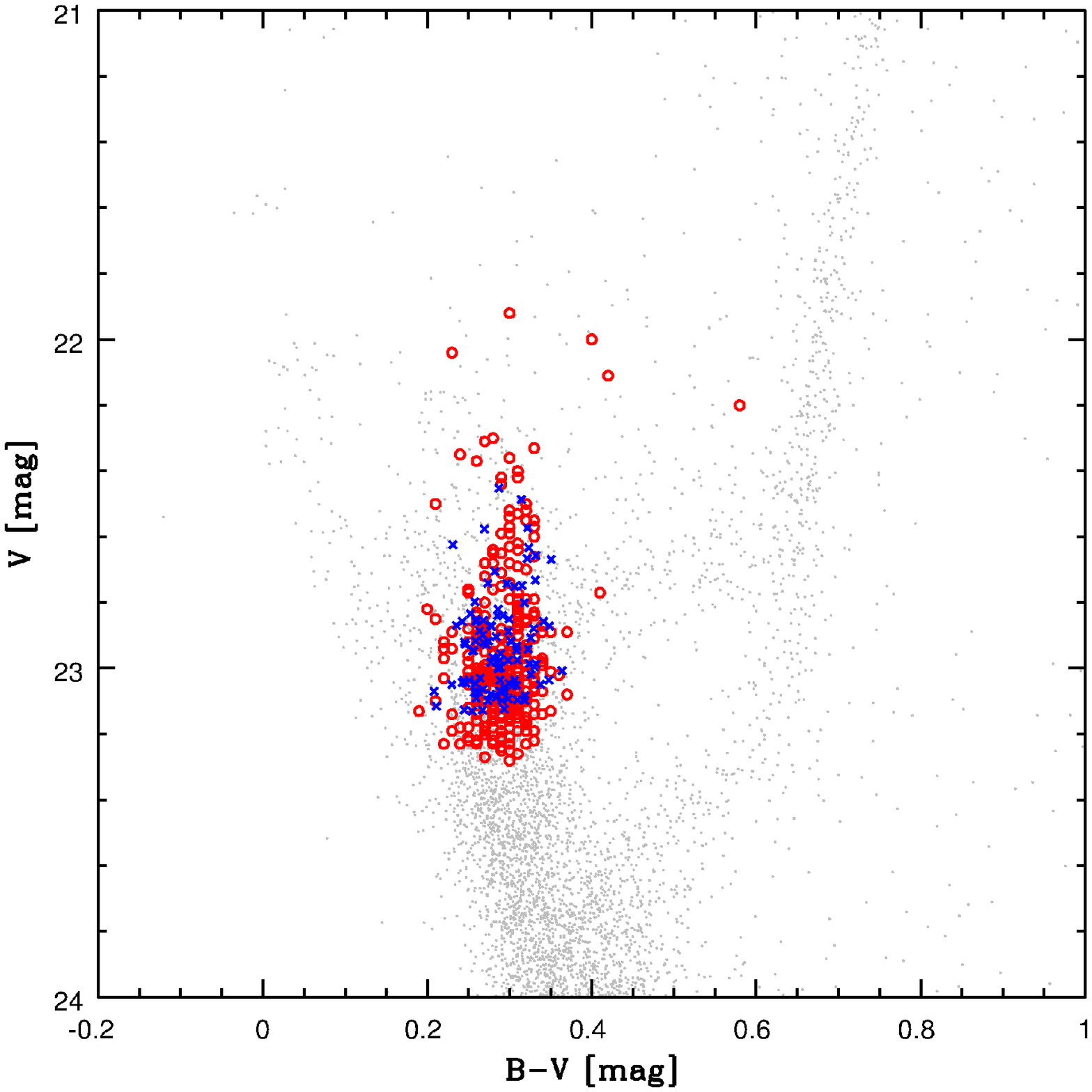} 
\caption{Position of SX Phe in $V$, $(B-V)$ CMD. Red circles and blue
  crosses display the VM13 and the new discovered variables, respectively. 
 their positions, periods and magnitudes  are
given in Tables~\ref{tab_sx_old} and~\ref{tab_sx_new}.}\label{cmd_sx} 
\end{figure*}

%%%%%%%%%%%%%%%%%%%%%%%%%%%%%%
\appendix
%%%%%%%%%%%%%%%%%%%%%%%%%%%%%%
\section{Notes on individual variables}

\textbf{V11, V26, V74}: Classified as a first-overtone pulsators by S86, according to the current 
photometry they seem to be double-mode pulsators.

\textbf{V17}: Classified as suspected variable in S86 and NV by D03, according to
the current photometry appears to be an EB with period 0.3933393
days. 

\textbf{V22}: Classified as first-overtone pulsator by S86 and D03. According to the current 
photometry it seems to pulsate in the fundamental mode with period
of days.  

\textbf{V25, V31}: Classified as first-overtone pulsator by S86. According to the current 
photometry it seems to pulsate in the fundamental mode with period
of days and they also show Blazhko effect.

\textbf{V27, V29, V33, V129 and V149}: Classified as fundamental mode pulsators by S86. According to the current 
photometry they appear to be anomalous cepheids.

\textbf{V40 and V115}: Classified as uncertain variables by S86,
according to our current set of data, they appear pulsate as
first-overtone and anomalous cepheid, respectively.

\textbf{V41 and V87}: Classified as suspected variables by S86,
according to our current set of data, they appear to be anomalous cepheids.

\textbf{V61, V77, V126, V127}: Classified as fundamental mode pulsators by S86. According to the current 
photometry they also show Blazhko effect.

\textbf{V89}: Classified as fundamental mode pulsator by S86. According to the current 
photometry it seems to be a double-mode pulsator. 

\textbf{V138 and V141}: Classified as suspected variables by S86,
according to our current set of data, they appear to be
fundamental-mode pulsators.

\textbf{V142 and V151}: Classified as suspected variables by S86,
according to our current set of data, they appear to be
first overtone pulsators.

\textbf{V176}: Classified as a first-overtone pulsators by D03
with period of $\sim 0.4$, respectively. According to the current 
photometry it seems to pulsate in the fundamental mode with period
of 0.764565 days. 

\textbf{V40, V65, V73, V84, V85, V123, V136, V138, V149, V177, V183, V186, V188, V195,
  V199, V201, V208, V211, V227, V228, V229, V230}: The $I$-band light 
curves are poorly sampled in the pulsation across maximum light.

\textbf{V31, V61, V76, V77, V126, V127, V206}: Classified as fundamental pulsators by
D03, according to the current photometry they seem to pulsate in the 
fundamental mode and they also show Blazhko effect.

\textbf{V74}: Classified as first overtone pulsator by
D03, according to the current photometry they seem to pulsate in the 
double mode.

\textbf{V85, V90, V181, V186, V196, V226}: The light curves are noisy.  

\textbf{V148}: This RR$_{c}$ is $\sim$0.3 mag brighter than the typical 
luminosity of HB stars.

\textbf{V158, V182}: We confirm the peculiar nature of the RRLs (see
also Paper VI).

\textbf{V161}: Classified as suspected variable in D03, according to the current photometry 
it does not show variability.

\textbf{V170 and V171}: Not enough data to fit a light curves, uncertain parameters.

\textbf{V176, V179, V196, V182, V200, V223}: These RRLs have very small
luminosity amplitudes for an ab-type RRL.

\textbf{V181}: This variable has very small luminosity amplitude for an c-type RRL. 
Possible blend with nearby faint star.

\textbf{V193, V205, V216 , V217, V219}: These variables have very
small amplitudes for an AC-type variable.
Possible blends with nearby faint stars.

\textbf{V3, V25, V44, V58 and V133}: These stars were classified as new discovered
variables in Paper VI with the identification numbers: V215, V213,
V212, V211 and V209, respectively.  

\textbf{V4, V32, V41 and V165}: These stars were classified in D03
and Paper VI as V177,
V202, V180 and V165, respectively.

\textbf{V218, V227, V228, V229, V230}: Poorly sampled light curves,
pulsational parameters are uncertain.

\textbf{RRL-1, RRL-2, RRL-3, RRL-4, RRL-5, RRL-38, AC-1, AC-9, AC-10}:
These stars are the six RRLs (three RR$_c$, three RR$_{ab}$) and the three ACs
recently detected by VM13 outside the tidal radius of Carina.

%%%%%%%%%%%%%%%%%%%%%%%%%%%%%%%%%%%%%%%%%%%%%%%%%%%%%%%%%%%%%%%%%%%%%%%%%%%%%%%%%%%%%%%%%%%%%%%%%%%%%%%%%%%

%==============================================================================================

\end{document}